\documentclass{jpp}
\usepackage{graphicx}
\usepackage[utf8]{inputenc}
\usepackage[T1]{fontenc}
\usepackage{amsmath}
\usepackage{mathtools}
\usepackage{bm}
\usepackage{booktabs}
\graphicspath{{figures/}}

\newcommand{\dd}{\mathrm{d}}
\newcommand{\ddt}{\mathrm{d}t}
\newcommand{\di}{d_i}
\newcommand{\vA}{v_A}
\newcommand{\Oci}{\Omega_{ci}}
\newcommand{\bhat}{\hat{\bm{b}}}
\newcommand{\UE}{\bm{U}_E}
\newcommand{\uperp}{\bm{u}_\perp}
\newcommand{\DE}{\mathrm{D}_E}
\newcommand{\Dg}{\mathrm{D}_g}
\newcommand{\Di}{\mathrm{D}_i}

\usepackage{xcolor}

\shorttitle{Eulerian ion energisation in HVM turbulence}
\shortauthor{L. Richard, S. S. Cerri, O. Pezzi, C. Granier and J. M. TenBarge}

\title{Eulerian ion energisation in three-dimensional\\ hybrid-Vlasov--Maxwell turbulence}

\author{
    Louis Richard\aff{1}\corresp{\email{louis.richard@physics.ox.ac.uk}},
    Silvio Sergio Cerri\aff{2},
    Oreste Pezzi\aff{3},
    Camille Granier\aff{4,5}
    \and Jason M. TenBarge\aff{6}
}

\affiliation{
    \aff{1} Rudolf Peierls Centre for Theoretical Physics, University of Oxford, Clarendon Laboratory, Parks Road, Oxford OX1 3PU, United Kingdom
    \aff{2} Université Côte d'Azur, Observatoire de la Côte d'Azur, CNRS, Laboratoire J.-L. Lagrange, Boulevard de l'Observatoire, CS 34229, 06304 Nice Cedex 4, France
    \aff{3} Istituto per la Scienza e Tecnologia dei Plasmi, Consiglio Nazionale delle Ricerche (ISTP-CNR), 70126 Bari, Italy
    \aff{4} Department of Physics, Stanford University, 382 Via Pueblo Mall, Stanford, CA 94305, USA
    \aff{5} Kavli Institute for Particle Astrophysics and Cosmology, Stanford University, Stanford, CA 94305, USA
    \aff{6} Department of Astrophysical Sciences, Princeton University, Princeton, NJ 08544, USA
}

\begin{document}

\maketitle

\begin{abstract}
    In collisionless plasma turbulence, energy is transferred from electromagnetic fluctuations to the particles.
    However, the relative contributions of different ion energisation mechanisms remain unclear.
    We investigate ion energisation in three-dimensional hybrid Vlasov--Maxwell simulations of decaying turbulence at $\beta_{i0}=0.25$ and $\beta_{i0}=4$, where $\beta_{i0}$ is the initial ratio of ion thermal to magnetic pressure.
    Using the exact ion momentum equation, we decompose the ion energisation rate $\bm{j}_i\bm{\cdot}\bm{E}$ into the work done on the parallel, diamagnetic, curvature, agyrotropic and polarisation ion currents.
    This decomposition requires neither gyrotropy nor scale separation and, therefore, remains valid where guiding-centre theory breaks down.
    We show that the agyrotropic term, which has no guiding-centre counterpart, accounts for $21\,\%$ of the energisation at $\beta_{i0}=0.25$ and $91\,\%$ at $\beta_{i0}=4$.
    We find that the dominant instantaneous energisation term changes with $\beta_{i0}$, from the polarisation current at low $\beta_{i0}$ to the diamagnetic and agyrotropic currents at high $\beta_{i0}$, at all resolved scales.
    In addition, we show that reversible exchange with the ion bulk flow dominates the instantaneous rate, so the time-integrated energy transfer is larger at $\beta_{i0}=4$ even though the instantaneous rate is smaller.
    Our results provide quantitative evidence that guiding-centre drifts alone cannot describe ion energisation in kinetic-scale turbulence and that the dominant energisation term depends on $\beta_i$.
    The results also highlight the key role of non-local thermodynamic-equilibrium effects in energy conversion in collisionless plasma turbulence.
\end{abstract}


\section{Introduction}\label{sec:intro}
Turbulence in weakly collisional, magnetised plasmas transfers energy from large-scale electromagnetic fluctuations to scales where it can be converted to particles~\citep{schekochihin_astrophysical_2009,matthaeus_turbulence_2021,cerri_kinetic_2019}.
Which kinetic mechanisms mediate that transfer, in what proportion, and how the partition depends on plasma parameters are long-standing questions central to the thermodynamics of the solar wind, planetary magnetosheaths, the intracluster medium, and accretion flows~\citep{howes_fundamental_2024}.

Much of the work to date has focused on individual energisation mechanisms and the rate at which each transfers field energy to particles.
Landau and transit-time damping, ion-cyclotron damping, stochastic heating, and energisation in reconnecting current sheets have each been studied in isolation and diagnosed directly in numerical simulations and in \textit{in situ} spacecraft observations~\citep{klein_measuring_2016,klein_diagnosing_2017,howes_diagnosing_2017,juno_field_2021,afshari_direct_2024,chandran_perpendicular_2010,hoppock_stochastic_2018,cerri_stochastic_2021,squire_highfrequency_2022,arzamasskiy_hybridkinetic_2019}.
A complementary line of work quantifies the exchange between bulk-flow and internal energy through the pressure--strain interaction $-(\bm{P}_s\bm{\cdot}\bm{\nabla})\bm{\cdot}\bm{U}_s$, where $\bm{U}_s$ and $\bm{P}_s$ are the bulk velocity and pressure tensor of species $s$, both in simulations and in spacecraft data~\citep{wan_intermittent_2015,yang_energy_2017,chasapis_situ_2018,bandyopadhyay_statistics_2020,pezzi_dissipation_2021,cassak_pressure_2022,barbhuiya_pressure_2022,conley_kinetic_2024}.
This approach can be extended beyond the second moment to the energy conversion carried by all higher moments of the distribution \citep{cassak_quantifying_2023} and to the generalised entropy production that accompanies it \citep{zhdankin_generalized_2022}, which separates the thermal from the non-thermal part of the velocity distribution function (VDF).
A gap remains between the two approaches.
The first attributes energisation to a basic plasma mechanism.
However, each mechanism must be identified and diagnosed separately, and no single one accounts for the whole budget.
The second closes the budget for the transfer between bulk and internal energy and, in its higher-moment form, resolves where that energy goes in velocity space, but it does not identify which field--particle interaction drove the transfer.
What is missing is an exact, complete, term-by-term decomposition of the work done by the fields that does not presuppose which mechanism operates.

Such a decomposition is obtained in the Eulerian frame by expressing the current of species $s$, $\bm{j}_s=q_sn_s\bm{U}_s$, through the generalised Ohm's law, that is, the momentum equation of the species solved for the electric field $\bm{E}$.
Here $q_s$, $n_s$ and $\bm{U}_s$ are the charge, number density and bulk velocity of species $s$.
The work $\bm{j}_s\bm{\cdot}\bm{E}$ then separates into the work done by $\bm{E}$ on the current parallel to the magnetic field $\bm{B}=B\hat{\bm{b}}$ and on the diamagnetic, curvature, agyrotropic and polarisation currents.
Because this approach follows from the exact momentum equation, it requires no ordering assumption and remains meaningful where guiding-centre theory fails.
\citet{tenbarge_electron_2024} introduced this Eulerian decomposition for electrons in magnetic reconnection \citep[see also][]{steinvall_eulerian_2025} and contrasted it with the Lagrangian, guiding-centre picture.
Its counterpart for ions in turbulence has not, to our knowledge, been examined.
For typical terrestrial magnetosheath conditions, the ordering parameter $\rho_i/L_B$ is routinely of order unity \citep{richard_nonmaxwellianity_2025}.
Here $L_B$ is the magnetic-field gradient scale and $\rho_i=v_{\mathrm{th}i,\perp}/\Oci$ is the ion gyroradius, with $v_{\mathrm{th},i\perp}=\sqrt{2T_{i\perp}/m_i}$ the ion thermal speed, $\Oci=eB/m_ic$ the ion cyclotron frequency, $T_{i\perp}$ the ion temperature perpendicular to $\hat{\bm{b}}$, $m_i$ the ion mass, $e$ the elementary charge and $c$ the speed of light.
As a result, the guiding-centre approach does not apply in such an environment, and an Eulerian description is required.

Two features of the ion energisation in turbulence make an Eulerian description necessary.
First, the natural adiabaticity parameter for ions is the ratio of the local magnetic-field radius of curvature to the gyroradius $r_c/\rho_i$ \citep{artemyev_rapid_2014,lemoine_particle_2023}.
The ratio falls to and below unity in the sheets and vortices that populate kinetic-scale turbulence, so that the ion VDF becomes agyrotropic~\citep{swisdak_quantifying_2016,servidio_kinetic_2015,valentini_differential_2016,yang_quantifying_2023,richard_nonmaxwellianity_2025}, and the magnetic moment $\mu=m_i v_\perp^2/2B$ is not conserved.
Here, $v_\perp$ is the particle velocity perpendicular to $\hat{\bm{b}}$.
Second, $\beta_i$ controls the relative size of the terms in Ohm's law.
At high $\beta_i$, ions decouple from the field over a broad range of scales, and the Hall and electron-pressure terms increasingly set the electric field~\citep{matteini_electric_2017,granier_electrononly_2024,califano_electrononly_2020}.
One therefore expects the partition of the budget among the currents, and not only its amplitude, to depend on $\beta_i$.

In this paper, we quantify the Eulerian ion energisation budget in three-dimensional, three-velocity-dimensional (3D3V) hybrid-Vlasov--Maxwell (HVM) simulations of decaying turbulence at $\beta_{i0}=0.25$ and $\beta_{i0}=4$.
Specifically, we address: (i) what can be said about ion energisation in strong turbulence when guiding-centre theory does not apply; (ii) which terms dominate the budget, and in particular whether those associated with the departure of the ion VDF from a gyrotropic Maxwellian matter; (iii) whether the local and volume-integrated pictures agree; (iv) how the budget evolves as the turbulence develops; and (v) how it depends on the plasma conditions.

The paper is organised as follows.
\S\ref{sec:equations} sets out the decomposition and its relation to the guiding-centre energisation terms, with the derivation given in Appendix~\ref{app:gc}.
\S\ref{sec:sims} describes the simulations.
\S\ref{sec:results} presents the spectra, ion VDF departures from a Maxwellian, magnetic-field geometry, and energisation budget.
\S\ref{sec:discussion} discusses the results and \S\ref{sec:conclusions} concludes.
Appendix~\ref{app:norm} gives the decomposition in the code normalisation.

\section{Ion energisation rate}\label{sec:equations}

\subsection{Eulerian decomposition}\label{ssec:eulerian}
We follow the derivation of \citet{tenbarge_electron_2024}, adapted to ions and written in the Gaussian units used by the HVM code.
The starting point is the exact ion momentum equation, obtained by taking the first moment of the Vlasov equation,
\begin{equation}
    \dfrac{\dd \bm{U}_i}{\ddt} + \dfrac{1}{m_i n_i} \bm{\nabla}\bm{\cdot}\bm{P}_i
    = \dfrac{e}{m_i}\left( \bm{E} + \dfrac{\bm{U}_i \times \bm{B}}{c}\right),
    \label{eq:momentum_balance}
\end{equation}
where $\dd/\ddt \equiv \partial_t + \bm{U}_i\bm{\cdot}\bm{\nabla}$ is the convective derivative following the bulk flow.
We write the pressure tensor as a gyrotropic part plus an agyrotropic part,
\begin{equation}
    \bm{P}_i = p_{i\perp}\,\bm{I}
             + \bigl(p_{i\parallel} - p_{i\perp}\bigr)\,\bhat\bhat
             + \bm{\Pi}_i^{\text{agy}},
    \label{eq:pressure_tensor_cgl}
\end{equation}
with $\bhat = \bm{B}/B$, $p_{i\parallel} = \bhat\bm{\cdot}\bm{P}_i\bm{\cdot}\bhat$ and $p_{i\perp} = \tfrac{1}{2}\left[\mathrm{Tr}(\bm{P}_i) - p_{i\parallel}\right]$.
The tensor $\bm{\Pi}_i^{\text{agy}}$ is defined by \eqref{eq:pressure_tensor_cgl}, so that it is traceless and carries all the non-gyrotropic information.

Substituting \eqref{eq:pressure_tensor_cgl} into \eqref{eq:momentum_balance}, solving for the electric field, and expanding the divergence of the anisotropy term with the magnetic curvature $\bm{\kappa}\equiv\bhat\bm{\cdot}\bm{\nabla}\bhat$, with local radius of curvature $r_c = |\bm{\kappa}|^{-1}$, gives
\begin{multline}
    \bm{E} + \dfrac{\bm{U}_i \times \bm{B}}{c}
    = \dfrac{1}{e n_i}\bm{\nabla} p_{i\perp}
    + \dfrac{\bhat}{e n_i}\bm{\nabla}\bm{\cdot}\!\bigl[\bigl(p_{i\parallel}-p_{i\perp}\bigr)\bhat\bigr]\\
    + \dfrac{p_{i\parallel}-p_{i\perp}}{e n_i}\,\bm{\kappa}
    + \dfrac{1}{e n_i}\bm{\nabla}\bm{\cdot}\bm{\Pi}_i^{\text{agy}}
    + \dfrac{m_i}{e}\dfrac{\dd\bm{U}_i}{\ddt}.
    \label{eq:ohm_expanded}
\end{multline}
Taking the cross product of \eqref{eq:ohm_expanded} with $\bm{B}$, using $(\bm{U}_i\times\bm{B})\times\bm{B}=-B^2\bm{U}_{i\perp}$ and multiplying by $c/B^2$ gives the perpendicular bulk velocity as a sum of drifts,
\begin{multline}
    \bm{U}_{i\perp}
    = \underbrace{\dfrac{c\,\bm{E}\times\bm{B}}{B^2}}_{\UE}
    + \underbrace{\dfrac{c}{e n_i B^2}\,\bm{B}\times\bm{\nabla} p_{i\perp}}_{\bm{U}_{i,\mathrm{dia}}}
    + \underbrace{\dfrac{c\bigl(p_{i\parallel}-p_{i\perp}\bigr)}{e n_i B^2}\,\bm{B}\times\bm{\kappa}}_{\bm{U}_{i,\mathrm{curv1}}}\\
    + \underbrace{\dfrac{c}{e n_i B^2}\,\bm{B}\times\bm{\nabla}\bm{\cdot}\bm{\Pi}_i^{\text{agy}}}_{\bm{U}_{i,\mathrm{agy}}}
    + \underbrace{\dfrac{m_i c}{e B^2}\,\bm{B}\times\dfrac{\dd\bm{U}_i}{\ddt}}_{\bm{U}_{i,\mathrm{pol}}}.
    \label{eq:Uperp}
\end{multline}
The second term on the right-hand side of \eqref{eq:ohm_expanded} is parallel to $\bhat$ and thus vanishes when taking the cross product with it.
%It contributes to the parallel dynamics only.
The last term of \eqref{eq:Uperp} is the inertial drift, which reduces to the familiar expression $(m_i c^2/eB^2)\,\partial_t\bm{E}_\perp$ for the polarisation drift in a uniform, static field, i.e., when $\bm{U}_i\simeq\UE$, and hence the name.
However, it is evaluated here with the full bulk acceleration. It therefore also contains the curvature drift of the bulk parallel flow, $m_i U_{i\parallel}^2 (c/eB^2)\,\bm{B}\times\bm{\kappa}$, and the advection of the remaining drifts by the flow (see Appendix~\ref{app:gc}).

The ion energisation rate follows from $\bm{j}_i\bm{\cdot}\bm{E} = e n_i \bm{U}_i\bm{\cdot}\bm{E} = j_{i\parallel} E_\parallel + e n_i \bm{U}_{i\perp}\bm{\cdot}\bm{E}_\perp$.
The $\UE$ term does no work, since $\UE\bm{\cdot}\bm{E}=0$ identically.
Each remaining term can be rearranged using the vector identity $(\bm{B}\times\bm{X})\bm{\cdot}\bm{E} = \bm{X}\bm{\cdot}(\bm{E}\times\bm{B})$ so that the $\bm{E}\times\bm{B}$ velocity appears explicitly,
\begin{multline}
    \bm{j}_{i}\bm{\cdot}\bm{E}
    = \underbrace{j_{i\parallel} E_\parallel}_{W_\parallel}
    + \underbrace{\UE\bm{\cdot}\bm{\nabla} p_{i\perp}}_{W_{\mathrm{dia}}}
    + \underbrace{\bigl(p_{i\parallel}-p_{i\perp}\bigr)\,\UE\bm{\cdot}\bm{\kappa}}_{W_{\mathrm{curv1}}}\\
    + \underbrace{\UE\bm{\cdot}\bm{\nabla}\bm{\cdot}\bm{\Pi}_i^{\text{agy}}}_{W_{\mathrm{agy}}}
    + \underbrace{m_i n_i\,\UE\bm{\cdot}\dfrac{\dd\bm{U}_i}{\ddt}}_{W_{\mathrm{pol}}}.
    \label{eq:J_dot_E}
\end{multline}
Equation~\eqref{eq:J_dot_E} is an exact identity; we make no approximations beyond the definitions of the pressure tensor in~\eqref{eq:pressure_tensor_cgl}, and in particular make no assumptions about gyrotropy, small $\rho_i/L_B$, or slow time variation.
Each term is the rate of work done by the electric field on one component of the ion current.
We therefore refer to them as \emph{terms} rather than as channels or mechanisms.

It will be useful below to regroup the diamagnetic and curvature terms by splitting $(p_{i\parallel}-p_{i\perp})\UE\bm{\cdot}\bm{\kappa}$ into $p_{i\parallel}\UE\bm{\cdot}\bm{\kappa} - p_{i\perp}\UE\bm{\cdot}\bm{\kappa}$,
\begin{equation}
    W_{\mathrm{dia}} + W_{\mathrm{curv1}}
    = \underbrace{\UE\bm{\cdot}\bm{\nabla} p_{i\perp}
      - p_{i\perp}\,\UE\bm{\cdot}\bm{\kappa}}_{W_{\mathrm{beta}}}
    + \underbrace{p_{i\parallel}\,\UE\bm{\cdot}\bm{\kappa}}_{W_{\mathrm{Fermi}}}.
    \label{eq:regrouping}
\end{equation}
The names are justified in \S\ref{sec:gc}.
Finally, expanding the polarisation term with $\bm{U}_i\bm{\cdot}\bm{\nabla}\bm{U}_i = \bm{\nabla}(U_i^2/2) - \bm{U}_i\times\bm{\omega}_i$, where $\bm{\omega}_i = \bm{\nabla}\times\bm{U}_i$ is the bulk vorticity, separates it into a local, a kinetic-energy-gradient and a vortical part,
\begin{equation}
    W_{\mathrm{pol}} = m_i n_i\,\UE\bm{\cdot}
    \left( \dfrac{\partial \bm{U}_i}{\partial t}
    + \bm{\nabla}\!\left(\dfrac{U_i^2}{2}\right) - \bm{U}_i \times \bm{\omega}_i\right).
    \label{eq:Wpol_vorticity}
\end{equation}

\subsection{Relation to guiding-centre energisation}\label{sec:gc}
The terms of \eqref{eq:J_dot_E} differ from the guiding-centre energisation terms.
In the Lagrangian guiding-centre (LGC) description, we obtain the ion energy evolution from the guiding-centre energy equation \citep{northrop_adiabatic_1963}, averaged over guiding centres in a local volume.
Previous studies of particle energisation retained the parallel electric field and the work done on the grad-$B$, curvature and polarisation drifts \citep{dahlin_mechanisms_2014,li_particle_2017}.
Following these studies, we obtain
\begin{equation}
    \dfrac{\dd \mathcal{E}_i^{\rm LGC}}{\ddt}
    = j_{i\parallel} E_\parallel
    + \underbrace{p_{i\perp}\,\DE \ln B}_{\text{betatron}}
    + \underbrace{\bigl(p_{i\parallel} + m_i n_i U_{i\parallel}^2\bigr)\,
      \UE\bm{\cdot}\bm{\kappa}}_{\text{Fermi}}
    + \underbrace{\dfrac{m_i n_i}{2}\,\Dg|\UE|^2}_{\text{polarisation}},
    \label{eq:gc_energisation}
\end{equation}
where $\DE \equiv \partial_t + \UE\bm{\cdot}\bm{\nabla}$ and $\Dg \equiv \partial_t + (\UE + U_{i\parallel}\bhat)\bm{\cdot}\bm{\nabla}$ are the convective derivatives along the $\bm{E}\times\bm{B}$ flow and the guiding centre, respectively, and $\mathcal{E}_i^{\rm LGC} = \tfrac{3}{2}p_i + \tfrac{1}{2}m_i n_i\bigl(U_{i\parallel}^2 + |\UE|^2\bigr)$.
The bulk-flow part of the Fermi term, $m_i n_i U_{i\parallel}^2\UE\bm{\cdot}\bm{\kappa}$, corresponds to the work done on the curvature drift of the bulk parallel flow, i.e., the centrifugal part.
In contrast to \eqref{eq:J_dot_E}, equation~\eqref{eq:gc_energisation} assumes that the magnetic moment is conserved and that $\rho_i/L_B\ll1$.

To relate the two descriptions, we use the exact evolution of the magnetic field strength along the $\bm{E}\times\bm{B}$ flow (see Appendix~\ref{app:gc}) and rewrite \eqref{eq:gc_energisation} in terms of the Eulerian terms of \eqref{eq:J_dot_E},
\begin{multline}
    \dfrac{\dd \mathcal{E}_i^{\rm LGC}}{\ddt}
    = W_\parallel + W_{\mathrm{dia}} + W_{\mathrm{curv1}}
    + \underbrace{m_i n_i U_{i\parallel}^2\,\UE\bm{\cdot}\bm{\kappa}
      + \dfrac{m_i n_i}{2}\,\Dg|\UE|^2}_{\textstyle W_{\mathrm{pol}}^{\rm LGC}}\\
    - \underbrace{\bm{\nabla}\bm{\cdot}\left(p_{i\perp}\UE\right)}_{\text{flux}}
    - \underbrace{\dfrac{\beta_{i\perp}}{2}\,J_\parallel E_\parallel}_{\text{twist}},
    \label{eq:gc_correspondence}
\end{multline}
where $\beta_{i\perp} = 8\pi p_{i\perp}/B^2$ and $J_\parallel$ is the \emph{total} parallel current.
Subtracting \eqref{eq:gc_correspondence} from \eqref{eq:J_dot_E}, we obtain the difference between the two descriptions,
\begin{equation}
    \bm{j}_i\bm{\cdot}\bm{E}
    - \dfrac{\dd \mathcal{E}_i^{\rm LGC}}{\ddt}
    = W_{\mathrm{agy}}
    + \bm{\nabla}\bm{\cdot}\left(p_{i\perp}\UE\right)
    + \dfrac{\beta_{i\perp}}{2}\,J_\parallel E_\parallel
    + W_{\mathrm{rot}}
    + W_{\perp},
    \label{eq:gc_residual}
\end{equation}
where $W_{\mathrm{rot}} = m_i n_i U_{i\parallel}\UE\bm{\cdot}\DE\bhat$ is the work associated with the rotation of the magnetic field direction along the $\bm{E}\times\bm{B}$ flow, and $W_\perp$ contains three terms of first and second order in the non-$\bm{E}\times\bm{B}$ perpendicular flow $\bm{u}_\perp=\bm{U}_{i\perp}-\UE$ (see Appendix~\ref{app:gc}).
We note that $W_{\mathrm{rot}}$ is part of the full guiding-centre energy equation, but previous studies neglected it \citep{dahlin_mechanisms_2014,li_particle_2017}.
Since $\bm{j}_i\bm{\cdot}\bm{E}$ is exact, \eqref{eq:gc_residual} captures the part of the work done by the field on the ions that the truncated guiding-centre description misses.

Equations~\eqref{eq:gc_correspondence} and~\eqref{eq:gc_residual} have several implications.
First, $W_{\mathrm{curv1}}$ does not correspond to the Fermi term.
$W_{\mathrm{curv1}}$ depends on the pressure \emph{anisotropy} $p_{i\parallel}-p_{i\perp}$, whereas the Fermi term depends on $p_{i\parallel}+m_in_iU_{i\parallel}^2$.
The difference $p_{i\perp}\UE\bm{\cdot}\bm{\kappa}$ contributes to the betatron term, while the centrifugal part is included in $W_{\mathrm{pol}}$.
Therefore, a small $W_{\mathrm{curv1}}$ does not imply weak Fermi energisation.
The rewriting in \eqref{eq:regrouping} separates the two contributions, i.e., $W_{\mathrm{beta}}$ corresponds to the betatron term up to the flux and twist terms, and $W_{\mathrm{Fermi}}$ to the Fermi term without its centrifugal part.
Second, the terms in \eqref{eq:gc_residual} are of two types.
On the one hand, the flux term integrates to zero over the periodic domain, and the twist term and $W_{\mathrm{rot}}$ remain finite even where guiding-centre theory is valid.
Thus, these terms emerge from the truncation underlying the usual derivation of \eqref{eq:gc_energisation}.
On the other hand, $W_{\mathrm{agy}}$ and $W_\perp$ vanish in the gyrotropic, small-$\rho_i/L_B$ limit and, therefore, quantify the breakdown of the guiding-centre description itself.
In particular, $W_{\mathrm{agy}}$ has no guiding-centre counterpart and provides the simplest diagnostic of this breakdown.
In this study, we use \eqref{eq:gc_residual} qualitatively (\S\ref{sec:discussion}) and leave a term-by-term evaluation for future work.

\subsection{Relation to the kinetic energisation mechanisms}
The terms in \eqref{eq:J_dot_E} also relate to the individual energisation mechanisms that transfer field energy to particles, as listed in \S\ref{sec:intro}.
The decomposition rearranges a fluid moment, whereas Landau and transit-time damping, cyclotron damping, and stochastic heating describe resonances in velocity space.
The two are therefore not in one-to-one correspondence, and each term collects both resonant and non-resonant contributions.
Broadly, ion Landau damping acts through $E_\parallel$ and is carried by $W_\parallel$~\citep{klein_diagnosing_2017}, while transit-time damping acts through the mirror force and enters through the betatron term, that is, through $W_{\mathrm{dia}}$ in \eqref{eq:J_dot_E}~\citep{arzamasskiy_hybridkinetic_2019}.
Ion-cyclotron damping is gyrophase-resonant, so it enters $W_{\mathrm{agy}}$ instantaneously, but its secular effect is the gyrotropic relaxation of the pressure anisotropy, which acts on the coefficients of $W_{\mathrm{curv1}}$ and $W_{\mathrm{dia}}$ although it does not do work through them \citep{kennel_velocity_1966,afshari_direct_2024}.
Stochastic heating~\citep{hoppock_stochastic_2018} and energisation in reconnecting current sheets~\citep{li_particle_2017} are not confined to a single term: the first appears among the perpendicular terms, and the second combines parallel acceleration with Fermi energisation.

\section{Numerical simulations}\label{sec:sims}

\subsection{Governing equations and numerical setup}\label{sec:setup}
We use the hybrid-Vlasov--Maxwell (HVM) code \citep{valentini_hybridvlasov_2007}, which advances the Vlasov equation for the ion phase-space density $f_i(\bm{x},\bm{v},t)$,
\begin{equation}
    \partial_{t} f_i + \bm{v}\bm{\cdot}\bm{\nabla} f_i
    + \dfrac{e}{m_i}
    \left(\bm{E} + \dfrac{\bm{v}}{c}\times\bm{B}\right)
    \bm{\cdot}\bm{\nabla}_{\bm{v}} f_i = 0.
    \label{eq:vlasov}
\end{equation}
The electrons are treated as a charge-neutralising fluid with a reduced mass ratio $m_i/m_e = 100$, so $d_e = 0.1\,\di$, where $d_i=c/\omega_{pi}=c/\left(4\pi n e^2 / m_i\right)^{1/2}$ is the ion inertial length.
The electric field is obtained from a generalised Ohm's law in which electron inertia is retained,
\begin{equation}
    \left(1 - d_e^{2}\nabla^{2}\right)\bm{E}
    = -\dfrac{\bm{U}_i}{c}\times\bm{B}
    + \dfrac{\bm{J}\times\bm{B}}{enc}
    - \dfrac{\bm{\nabla} P_e}{en}
    + \dfrac{4\pi d_e^{2}}{c^{2}}\bm{\nabla}
    \bm{\cdot}\left(\bm{U}_i\bm{J} + \bm{J}\bm{U}_i
    - \dfrac{\bm{J}\bm{J}}{en}\right),
    \label{eq:ohms_law}
\end{equation}
\noindent
where $\bm{J} = (c/4\pi)\bm{\nabla}\times\bm{B}$ is the total current density neglecting the displacement current.
Equation~\eqref{eq:ohms_law} also assumes quasi-neutrality so that $\partial_t \bm{J} \simeq (c^2/4\pi)\nabla^2 \bm{E}$.
We use an isothermal electron closure $P_e=nT_{e0}$ and $\beta_{e0}=8\pi p_{e0}/B^2=0.1$ in both runs.

Table~\ref{tab:runs} summarises the parameters of the two runs.
These runs were previously used to study electron-only reconnection in decaying turbulence~\citep{granier_electrononly_2024}.
The spatial domain is a triply periodic cube of side $L=6\pi\,\di$ discretised on a uniform $N_p^3$ grid, while velocity space spans $[-U, U]$ in each direction on an $N_v^3$ grid.
With $N_p=256$, the grid spacing is $\Delta x = 0.074\,\di = 0.74\,d_e$, so that the electron inertial length is resolved by fewer than two cells and the sub-$d_e$ range is not resolved.
High-order compact filters \citep{lele_compact_1992} suppress spurious energy accumulation at the grid scale.

\begin{table}
    \begin{center}
        \def~{\hphantom{0}}
        \begin{tabular}{lccccccccc}
            & & & & \multicolumn{3}{c}{Position space} & \multicolumn{3}{c}{Velocity space}\\
            \cmidrule(lr){5-7} \cmidrule(lr){8-10}
            & $\beta_{i0}$ & $\delta B_{\mathrm{rms},0}/B_0$ & $\beta_e$
            & Dim. & $L~(\di)$ & $N_p$ & Dim. & $U~(v_{\mathrm{th},i})$ & $N_v$ \\[2pt]
            Run A & 0.25 & 0.5 & 0.1 & 3D & $6\pi$ & 256 & 3V & 7.0 & 57 \\
            Run B & 4~~~ & 0.5 & 0.1 & 3D & $6\pi$ & 256 & 3V & 5.0 & 57 \\
        \end{tabular}
        \caption{Parameters of the two hybrid-Vlasov--Maxwell runs.
        Both start from isotropic ion VDFs and electron pressure, $T_{i\perp}/T_{i\parallel}=T_{e\perp}/T_{e\parallel}=1$, with magnetic fluctuations seeded in $0.3\le k\di\le1$.}
        \label{tab:runs}
    \end{center}
\end{table}

At $t\Oci=0$, a uniform background field $\bm{B}_0 = B_0\hat{\bm{z}}$ is imposed.
Superposed on it are isotropic three-dimensional magnetic fluctuations $\delta\bm{B} = \delta\bm{B}_\perp + \delta B_\parallel\hat{\bm{z}}$, initialised through a vector potential perturbation $\delta\bm{A}$ with $\delta\bm{B} = \bm{\nabla}\times\delta\bm{A}$.
The fluctuations are seeded in the narrow band $0.3 \leq k\di \leq 1$ with root-mean-square amplitude $\delta B_{\mathrm{rms},0}/B_0 = 0.5$.
% NOTE: please state whether velocity fluctuations are also initialised.
The turbulence is therefore injected near the ion scales, which is relevant to planetary magnetosheaths such as Earth's, where the correlation length is typically $\lambda_c\sim 10\,\di$~\citep{stawarz_properties_2019}, rather than the solar wind, where an extended MHD cascade precedes the ion scales.

Unless stated otherwise, all quantities are shown at the time of peak root-mean-square current density, $J_{\mathrm{rms}}=\langle|\bm{J}-\langle\bm{J}\rangle|^2\rangle^{1/2}$.
This peak occurs at $t\Oci = 13.20$ for $\beta_{i0}=0.25$ and $t\Oci = 15.60$ for $\beta_{i0}=4$, and it is the conventional proxy for the time at which the turbulence is fully developed \citep{servidio_magnetic_2009,wan_intermittent_2015}.

\subsection{Diagnostics}\label{sec:diagnostics}
We compute the moments $n_i$, $\bm{U}_i$, and $\bm{P}_i$ directly from $f_i$ on the velocity grid, and evaluate spatial derivatives spectrally, as the solver does \citep{valentini_hybridvlasov_2007}.

The convective derivative in $\bm{U}_{i,\mathrm{pol}}$ and $W_{\mathrm{pol}}$ requires $\partial_t\bm{U}_i$.
However, the field dumps are separated by $\Delta t\,\Oci \simeq 1.2$, which is too coarse to resolve it.
Therefore, we evaluate the polarisation drift and the polarisation term as residuals,
\begin{subequations}
    \begin{align}
        \bm{U}_{i,\mathrm{pol}} &= \bm{U}_{i\perp} - \UE - \bm{U}_{i,\mathrm{dia}} - \bm{U}_{i,\mathrm{curv1}} - \bm{U}_{i,\mathrm{agy}},
        \label{eq:Upol_residual} \\
        W_{\mathrm{pol}} &= e n_i\,\bm{U}_{i,\mathrm{pol}}\bm{\cdot}\bm{E}
        = \bm{j}_i\bm{\cdot}\bm{E} - W_\parallel - W_{\mathrm{dia}} - W_{\mathrm{curv1}} - W_{\mathrm{agy}}.
        \label{eq:Wpol_residual}
    \end{align}
\end{subequations}
This is not an approximation.
Substituting the momentum equation \eqref{eq:momentum_balance} for $\dd\bm{U}_i/\ddt$ in $W_{\mathrm{pol}}$ and using $\UE\bm{\cdot}\bm{E}=0$ gives \eqref{eq:Wpol_residual} identically.
However, the computed $W_{\mathrm{pol}}$ includes discretisation errors in the other four terms and numerical dissipation.
Since the budget closes by construction, its closure does not test the accuracy of the calculation.

\section{Results}\label{sec:results}

\subsection{Development of the turbulence and fluctuation spectra}\label{sec:spectra}
\begin{figure}
    \centering
    \includegraphics[width=\linewidth]{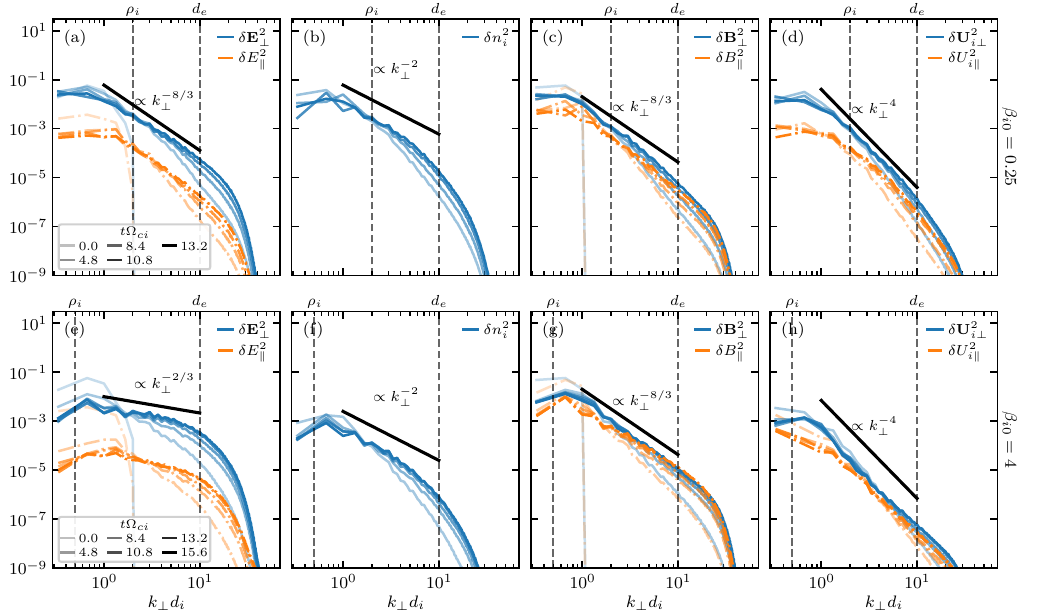}
    \caption{Time evolution of the reduced perpendicular power spectra for $\beta_{i0}=0.25$ (top row) and $\beta_{i0}=4$ (bottom row).
    (a,e) Electric field, (b,f) ion density, (c,g) magnetic field and (d,h) ion bulk velocity.
    Solid and dot-dashed lines show the perpendicular and parallel components, respectively.
    Line darkness increases with time, as indicated in the legend.
    Vertical dashed lines mark $k_\perp\rho_i=1$ and $k_\perp d_e=1$, and black solid lines indicate the reference power laws quoted in the text.
    All quantities are in code units.}
    \label{fig:spectra}
\end{figure}

Figure~\ref{fig:spectra} shows the time evolution of the reduced perpendicular power spectra $P(k_\perp)$ of the electric field, ion density, magnetic field, and ion bulk velocity for runs A and B.
Throughout the paper, we define $k_\perp$ with respect to the background magnetic field $\bm{B}_0$, not the local field, so $k_\perp=\sqrt{k_x^2 + k_y^2}$.
For $\beta_{i0}=0.25$ (figure~\ref{fig:spectra}a--d), the electric and magnetic spectra are close to one another around the ion gyroradius.
Both are consistent with a $k_\perp^{-8/3}$ scaling over $1\lesssim k_\perp\di\lesssim5$, as expected for a kinetic-Alfvén cascade~\citep{boldyrev_spectrum_2012}.
%At $k_\perp\di\gtrsim 6$, the two separate, the magnetic spectrum steepening while the electric spectrum flattens.
At $k_\perp d_e\gtrsim 1$, both fall off exponentially under the numerical filter acting at the grid scale.
The electric-field spectrum between $\rho_i$ and $d_e$ is somewhat steeper than the $P_E(k_\perp)\sim k_\perp^2P_B(k_\perp)$ expected when the Hall term sets $\bm{E}_\perp$.
We attribute this to the short sub-ion range $\rho_i=0.5\,d_i=5\,d_e$, and the numerical filter acts from $k_\perp d_e\simeq1$.
Decomposing $\bm{E}$ into the terms of \eqref{eq:ohms_law} (not shown) confirms that the Hall term dominates below $\rho_i$ and follows $k_\perp^2P_B$ just below $\rho_i$ before steepening.

For $\beta_{i0}=4$ (figure~\ref{fig:spectra}e--h), the magnetic spectrum is nearly unchanged, but the electric spectrum is shallower, close to $k_\perp^{-2/3}$.
This follows from the change of regime in Ohm's law.
When the ions decouple from the magnetic field, the perpendicular electric field is set not by the ideal term $-\bm{U}_i\times\bm{B}/c$ but by the Hall and electron-pressure terms.
Both scale as $k_\perp\delta B$, so that $P_E(k_\perp) \sim k_\perp^{2}P_B(k_\perp) \sim k_\perp^{-2/3}$.
The same flattening is observed in the Earth's magnetosheath \citep{matteini_electric_2017,chen_nature_2017}.

In both runs, the ion bulk velocity spectrum steepens sharply below the ion gyroradius, with an effective slope close to $-4$.
This is consistent with the strong decoupling of the ion flow from the magnetic fluctuations reported in hybrid and fully kinetic 3D simulations~\citep{franci_solar_2018,groselj_fully_2018,cerri_kinetic_2019}.

The spectra therefore indicate that run A resolves the $k_\perp\rho_i<1$ physics well but has only a short sub-ion range, limited by the reduced mass ratio.
In contrast, run B displays a fully developed sub-ion cascade, with more than one decade separating $\rho_i=2\di$ from $d_e=0.1\di$.

\subsection{Departures of the ion VDFs from a Maxwellian}\label{sec:nonmax}

Two terms of \eqref{eq:J_dot_E} require the ion VDF to depart from an isotropic Maxwellian.
$W_{\mathrm{curv1}}$ vanishes unless the pressure is anisotropic \emph{and} the field is curved along the $\bm{E}\times\bm{B}$ flow, while $W_{\mathrm{agy}}$ vanishes unless the ion VDF is non-gyrotropic.
We therefore characterise those departures before turning to the field geometry and the budget.

\begin{figure}
    \centering
    \includegraphics[width=\linewidth]{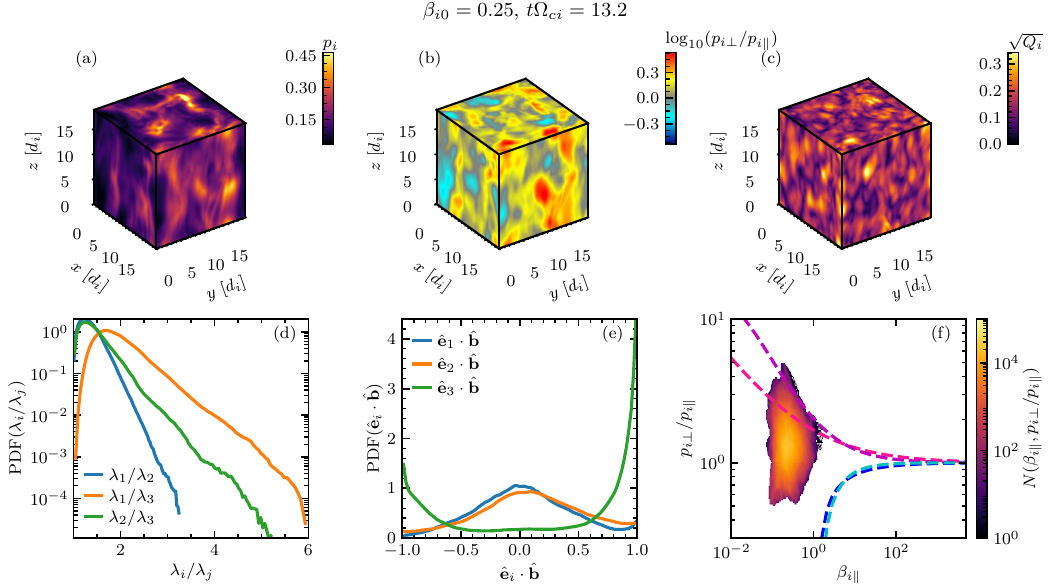}
    \caption{Departures of the ion VDFs from an isotropic Maxwellian for run A ($\beta_{i0}=0.25$) at $t\Oci=13.20$.
    (a) Ion scalar pressure $p_i$, (b) pressure anisotropy $\log_{10}(p_{i\perp}/p_{i\parallel})$ and (c) agyrotropy $\sqrt{Q_i}$, all shown on the faces of the simulation domain.
    (d) Distributions of the ratios of the eigenvalues of the ion pressure tensor.
    (e) Distributions of the alignment of the corresponding principal directions with the local magnetic field.
    (f) Joint distribution of $p_{i\perp}/p_{i\parallel}$ and $\beta_{i\parallel}$, with the mirror, ion-cyclotron, parallel- and oblique-firehose thresholds (for $\gamma=10^{-2}\Oci$) shown as dashed lines.}
    \label{fig:nonmax_lowbeta}
\end{figure}

Figure~\ref{fig:nonmax_lowbeta} shows, for run A, the isotropic ion pressure $p_i=\tfrac{1}{3}\mathrm{Tr}(\bm{P}_i)$, the pressure anisotropy $p_{i\perp}/p_{i\parallel}$ and Swisdak's agyrotropy measure \citep{swisdak_quantifying_2016}
\begin{equation}
    \sqrt{Q_i} = \left(\dfrac{p_{i,12}^2 + p_{i,13}^2 + p_{i,23}^2}
                  {p_{i\perp}^2 + 2p_{i\parallel} p_{i\perp}}\right)^{1/2}
    \in [0,1],
    \label{eq:swisdak}
\end{equation}
where the off-diagonal elements are taken in the local field-aligned frame oriented such that the last two diagonal components are equal.

The pressure anisotropy spans a broad range, $0.5 \lesssim p_{i\perp}/p_{i\parallel} \lesssim 5$.
The joint distribution of $p_{i\perp}/p_{i\parallel}$ and $\beta_{i\parallel}=8\pi p_{i\parallel}/B^2$ (figure~\ref{fig:nonmax_lowbeta}f) has its upper envelope bounded by the mirror-instability threshold \citep{hellinger_solar_2006}, as previously reported for turbulent plasmas in the Earth's magnetosheath~\citep{richard_nonmaxwellianity_2025}.
Nevertheless, the bulk of the distribution lies well inside that bound, near isotropy.
Measured against the empirical threshold in figure~\ref{fig:nonmax_lowbeta}(f), which corresponds to a growth rate $\gamma=10^{-2}\Oci$, only $0.06\,\%$ of the grid cells are potentially mirror unstable.
The lower envelope, $p_{i\perp}/p_{i\parallel}\simeq0.5$, is not set by the firehose thresholds, which are non-positive at these $\beta_{i\parallel}$ and therefore unreachable.

The agyrotropy locally reaches $\sqrt{Q_i}\simeq0.42$, although its volume average is only $0.14$.
We decompose the ion pressure tensor into its principal axes, with eigenvalues $\lambda_1\ge\lambda_2\ge\lambda_3$ and eigenvectors $\hat{\bm{e}}_1,\hat{\bm{e}}_2,\hat{\bm{e}}_3$.
We find (figure~\ref{fig:nonmax_lowbeta}e) that $\hat{\bm{e}}_3$ is preferentially aligned or anti-aligned with the local magnetic field $\hat{\bm{b}}$, which indicates that the ion VDF is narrowest along $\hat{\bm{b}}$, i.e., a strong predominance of perpendicular-heating processes in low-beta turbulence.
This result is consistent with the heavy tails of the $\lambda_1/\lambda_3$ and $\lambda_2/\lambda_3$ distributions (figure~\ref{fig:nonmax_lowbeta}d) and with the skew of $p_{i\perp}/p_{i\parallel}$ towards values greater than unity.

\begin{figure}
    \centering
    \includegraphics[width=\linewidth]{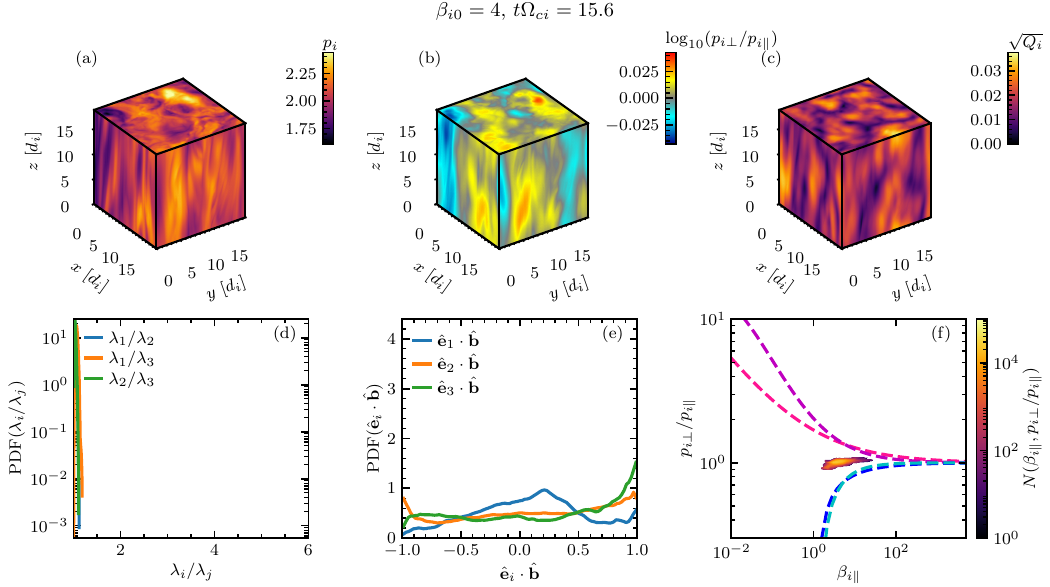}
    \caption{Departures of the ion VDFs from an isotropic Maxwellian for run B ($\beta_{i0}=4$) at $t\Oci=15.60$.
    (a) Ion scalar pressure $p_i$, (b) pressure anisotropy $\log_{10}(p_{i\perp}/p_{i\parallel})$ and (c) agyrotropy $\sqrt{Q_i}$, all shown on the faces of the simulation domain.
    (d) Distributions of the ratios of the eigenvalues of the ion pressure tensor.
    (e) Distributions of the alignment of the corresponding principal directions with the local magnetic field.
    (f) Joint distribution of $p_{i\perp}/p_{i\parallel}$ and $\beta_{i\parallel}$, with the mirror (magenta), ion-cyclotron (pink), parallel-firehose (blue) and oblique-firehose (cyan) thresholds (for $\gamma=10^{-2}\Oci$) shown as dashed lines.}
    \label{fig:nonmax_highbeta}
\end{figure}

Figure~\ref {fig:nonmax_highbeta} shows the same analysis for run B.
Consistent with the spectra, the ion pressure develops much finer sub-ion-scale structure than at low $\beta_{i0}$.
Both the pressure anisotropy and the agyrotropy are strongly reduced, $0.9 \lesssim p_{i\perp}/p_{i\parallel} \lesssim 1.1$ and $\sqrt{Q_i}\lesssim0.05$, an order of magnitude smaller than in run A.
In addition, the whole anisotropy distribution lies well inside all four thresholds.
The sub-ion-scale turbulence therefore does not drive significant departures from a gyrotropic Maxwellian at this $\beta_{i0}$, and no instability can realistically constrain the anisotropy.
Because the ion VDFs are nearly isotropic, the principal axes are nearly degenerate, and their orientation relative to $\hat{\bm{b}}$ is nearly random.

In summary, at $\beta_{i0}=0.25$ the ion VDFs develop a broad range of pressure anisotropies whose upper envelope follows the mirror threshold, together with substantial local agyrotropy.
At $\beta_{i0}=4$, they remain close to isotropic and well inside every threshold.
In neither run is the bulk of the distribution held at a marginal-stability boundary.
We return in \S\ref{sec:discussion} to the apparent paradox that $W_{\mathrm{agy}}$ nevertheless dominates the budget at $\beta_{i0}=4$, where the agyrotropy is smallest.

\subsection{Magnetic-field geometry}\label{sec:curvature}
\begin{figure}
    \centering
    \includegraphics[width=\linewidth]{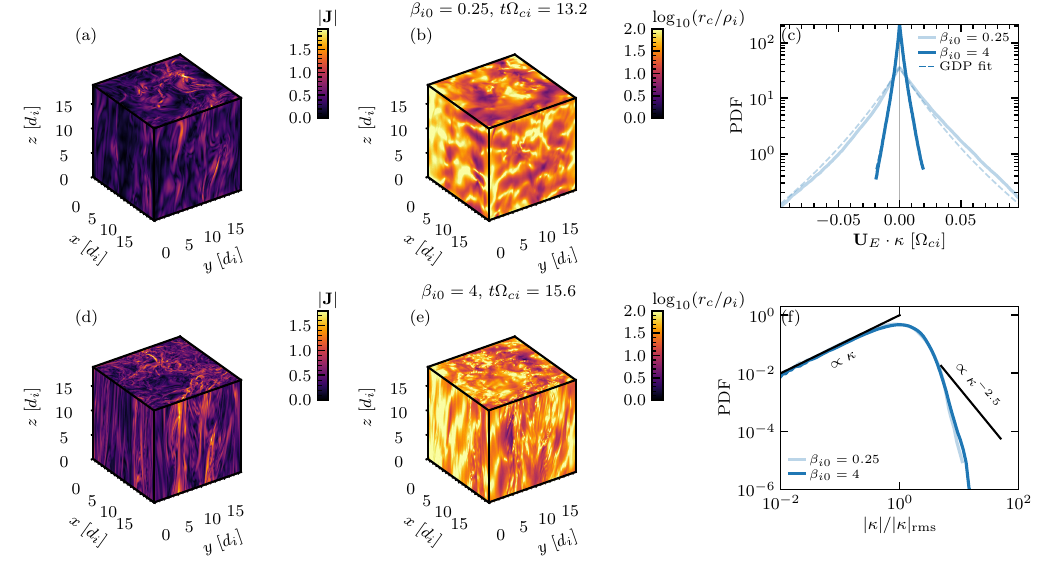}
    \caption{Magnetic-field geometry for run A ($\beta_{i0}=0.25$, $t\Oci=13.20$, top row) and run B ($\beta_{i0}=4$, $t\Oci=15.60$, bottom row).
    (a,d) Current density magnitude $|\bm{J}|$ and (b,e) $\log_{10}(r_c/\rho_i)$, the radius of curvature $r_c=|\bm{\kappa}|^{-1}$ normalised to the local ion gyroradius, both on the faces of the simulation domain.
    (c) Probability density function of $\UE\bm{\cdot}\bm{\kappa}$, the rate common to the curvature and Fermi terms, on a linear abscissa.
    The dashed lines are fits to \eqref{eq:gdp}. (f) Probability density function of $|\bm{\kappa}|$ normalised to its root-mean-square value, with the $|\bm{\kappa}|$ and $|\bm{\kappa}|^{-2.5}$ reference slopes indicated.
    In (c) and (f), both runs are shown, distinguished by opacity.
    All quantities are in code units.}
    \label{fig:curvature}
\end{figure}

The magnetic-field curvature enters ion energisation through $\UE\bm{\cdot}\bm{\kappa}$.
In addition, the loss of $\mu$ conservation in strongly curved regions ($r_c\sim \rho_i$) renders the ion motion chaotic and leads to pitch-angle scattering~\citep{artemyev_rapid_2014,lemoine_particle_2023,richard_nonmaxwellianity_2025}.
Figure~\ref{fig:curvature} shows, for both runs, the current density magnitude and the radius of curvature normalised to the local ion gyroradius, $r_c/\rho_i$, together with the distributions of $\UE\bm{\cdot}\bm{\kappa}$ and of $|\bm{\kappa}|$ normalised to its root-mean-square value $|\bm{\kappa}|_{\mathrm{rms}}$.
The gyroradius is evaluated cell by cell as $\rho_i = \sqrt{\beta_{i\perp}}\,\di$, with $\beta_{i\perp}=8\pi p_{i\perp}/B^2$ the local perpendicular ion beta.
% NOTE: state whether the rms products below use the local or the nominal rho_i.
The regions of strongest curvature are organised into thin, sheet-like structures that coincide with current sheets.
In both runs, $r_c$ falls to the local gyroradius there, $r_c/\rho_i \lesssim 1$, while the bulk of the volume has $r_c \gtrsim 10\,\rho_i$.
The magnetic moment is therefore not an adiabatic invariant in the regions expected to dominate energisation \citep{wan_intermittent_2015}, which justifies the Eulerian approach adopted here.

The distribution of $|\bm{\kappa}|/|\bm{\kappa}|_{\mathrm{rms}}$ is insensitive to $\beta_{i0}$.
In both runs, it rises as $|\bm{\kappa}|$ at small curvature, as required for a three-dimensional isotropic random field~\citep{yang_role_2019,bandyopadhyay_situ_2020a}, and peaks at the rms value, which is $|\bm{\kappa}|_{\mathrm{rms}}=0.1\,\di^{-1}\approx 0.2 \,\ell_\perp^{-1}$ at $\beta_{i0}=0.25$ and $0.03\,\di^{-1}\approx0.05 \,\ell_\perp^{-1}$ at $\beta_{i0}=4$, where $\ell_\perp$ is the perpendicular correlation length of the magnetic field.
Measured against the gyroradius, the two runs are almost indistinguishable, $|\bm{\kappa}|_{\mathrm{rms}}\,\rho_i = 0.05$ at $\beta_{i0}=0.25$ and $0.06$ at $\beta_{i0}=4$, so that $|\bm{\kappa}|_{\mathrm{rms}}\propto\rho_i^{-1}$.
In neither run does the tail reach the $|\bm{\kappa}|^{-2.5}$ scaling predicted for MHD turbulence \citep{yang_role_2019} and reported in the magnetosheath~\citep{bandyopadhyay_situ_2020a,richard_nonmaxwellianity_2025}.
Both fall off considerably faster.
The same behaviour was found in the fully kinetic simulations of \citet{sebastian_magnetic_2025} and follows from the moderate fluctuation amplitude, $\delta B_{\mathrm{rms},0}/B_0=0.5$, which is too weak to produce sharp field-line bends.

The distribution of $\UE\bm{\cdot}\bm{\kappa}$ (figure~\ref{fig:curvature}c) is sharply peaked at zero with near-exponential tails.
It is close to symmetric at $\beta_{i0}=0.25$ (skewness $0.04$) but appreciably skewed at $\beta_{i0}=4$ (skewness $0.86$).
Its width differs strongly between the runs, $|\UE\bm{\cdot}\bm{\kappa}|_{\text{rms}}=0.024\,\Oci$ at $\beta_{i0}=0.25$ against $0.004\,\Oci$ at $\beta_{i0}=4$.
This behaviour reflects an order-of-magnitude decrease in ion-velocity fluctuations (figure~\ref{fig:spectra}d,h), a threefold decrease in rms curvature, and weaker alignment between $\UE$ and $\bm{\kappa}$.
Alfvénic fluctuations are expected to result in a near-symmetric distribution, with the flow and curvature out of phase so that field-line contraction and expansion nearly cancel~\citep{sebastian_magnetic_2025}.
The positive skewness at $\beta_{i0}=4$ indicates a small net excess of contraction.

\subsection{Spatial distribution and volume-integrated budget}\label{sec:budget}

\begin{figure}
    \centering
    \includegraphics[width=\linewidth]{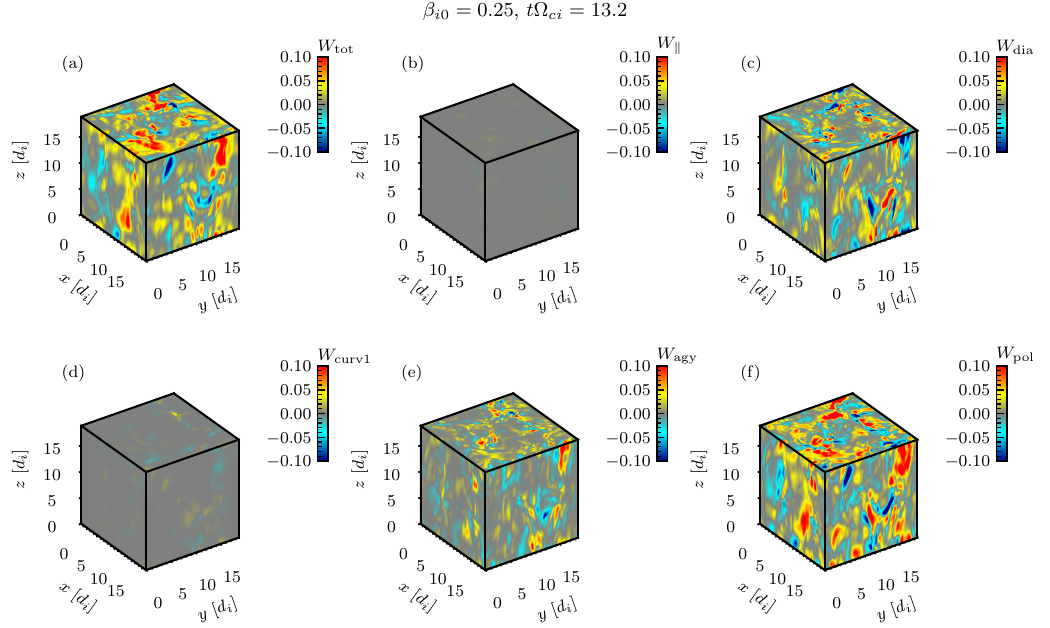}
    \caption{Decomposition of the ion energisation rate for run A ($\beta_{i0}=0.25$) at $t\Oci=13.20$.
    (a) Total $W_{\mathrm{tot}}=\bm{j}_i\bm{\cdot}\bm{E}$, (b) parallel $W_\parallel$, (c) diamagnetic $W_{\mathrm{dia}}$, (d) curvature $W_{\mathrm{curv1}}$, (e) agyrotropic $W_{\mathrm{agy}}$ and (f) polarisation $W_{\mathrm{pol}}$, all on the same colour scale and in code units.}
    \label{fig:ws-beta0.25}
\end{figure}

\begin{figure}
    \centering
    \includegraphics[width=\linewidth]{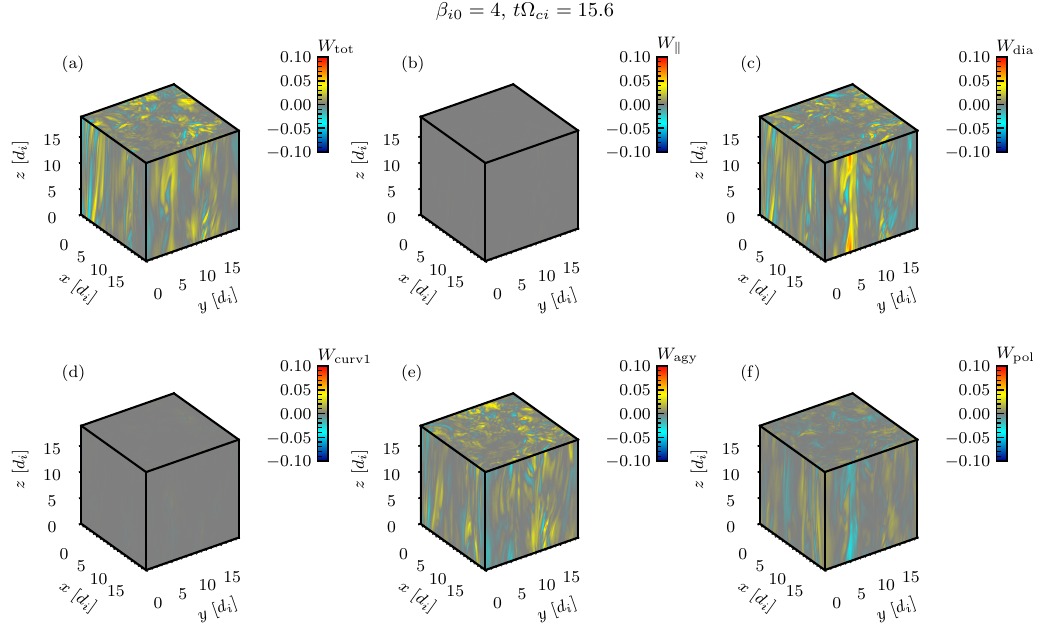}
    \caption{Decomposition of the ion energisation rate for run B ($\beta_{i0}=4$) at $t\Oci=15.60$.
    (a) Total $W_{\mathrm{tot}}=\bm{j}_i\bm{\cdot}\bm{E}$, (b) parallel $W_\parallel$, (c) diamagnetic $W_{\mathrm{dia}}$, (d) curvature $W_{\mathrm{curv1}}$, (e) agyrotropic $W_{\mathrm{agy}}$ and (f) polarisation $W_{\mathrm{pol}}$, all on the same colour scale and in code units.}
    \label{fig:ws-beta4}
\end{figure}

Figures~\ref{fig:ws-beta0.25} and~\ref{fig:ws-beta4} show the total ion energisation $W_{\mathrm{tot}} = \bm{j}_i\bm{\cdot}\bm{E}$ and each term of \eqref{eq:J_dot_E} for the two runs, on a common colour scale.
The energisation is strongly intermittent.
It is concentrated in thin structures that occupy a small fraction of the volume and alternate in sign, as is generic in kinetic turbulence~\citep{wan_intermittent_2015,chasapis_situ_2018,bandyopadhyay_statistics_2020}.
The parallel and curvature terms are visually negligible in both runs.
In contrast, the diamagnetic, agyrotropic and polarisation terms reach comparable amplitudes, with $W_{\mathrm{pol}}$ organised into the largest-scale structures and $W_{\mathrm{agy}}$ into the finest.

\vfill

Table~\ref{tab:budget} gives the volume-integrated budget.
At $\beta_{i0}=0.25$, the polarisation current dominates, carrying $76\,\%$ of the total, followed by the agyrotropic term at $21\,\%$.
The diamagnetic term contributes only $5\,\%$, and the curvature and parallel terms are at the per cent level, with the former opposite in sign to the total.
At $\beta_{i0}=4$, the composition changes qualitatively.
The total is $3.9$ times smaller; the agyrotropic term alone carries $91\,\%$, and the diamagnetic term $45\,\%$, while the polarisation term reverses sign and removes $34\,\%$.
We return to the sign reversal in \S\S\ref{sec:time} and \ref{sec:beta}
The parallel and curvature terms are negligible in both runs.
The last three columns of table~\ref{tab:budget} give the pressure--strain interaction and its two parts.
They are not terms of \eqref{eq:J_dot_E} but measure a different exchange, and we return to them in \S\ref{sec:discussion}.

\begin{figure}
    \centering
    \includegraphics[width=\linewidth]{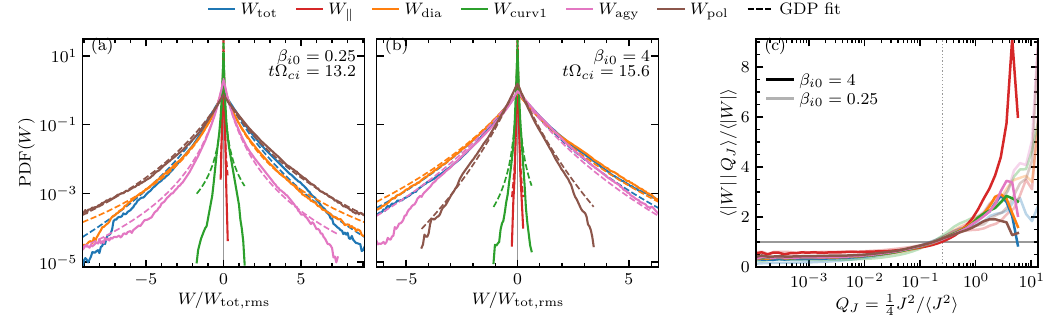}
    \caption{Probability density functions of the terms of \eqref{eq:J_dot_E} normalised to the rms total ion energisation $W_{\mathrm{tot},\mathrm{rms}}$ for (a) run A ($\beta_{i0}=0.25$) and (b) run B ($\beta_{i0}=4$) at the time of peak $J_{\mathrm{rms}}$.
    Solid lines show the measured distributions, and dashed lines show the generalised double Pareto fits \eqref{eq:gdp}, with parameters listed in table~\ref{tab:budget}.
    (c) Average of $|W|$ conditioned on $Q_J=\tfrac{1}{4}\bm{J}^2/\langle\bm{J}^2\rangle$ and normalised to the unconditional average, for both runs (distinguished by opacity).
    The horizontal line marks unity, and the vertical dotted line marks the mean $\langle Q_J\rangle=1/4$.
    All quantities are in code units.}
    \label{fig:ws-pdfs}
\end{figure}

Figure~\ref{fig:ws-pdfs} shows the corresponding distributions of the ion energisation terms normalised to the rms total energisation $W_{\mathrm{tot,rms}}$.
Similar to the distribution of $\UE\bm{\cdot}\bm{\kappa}$, they are sharply peaked at zero with pronounced non-Gaussian wings.
The volume integrals in table~\ref{tab:budget} therefore accumulate in a small fraction of cells.
We characterise these distributions by fitting a generalised double Pareto (GDP) form,
\begin{equation}
    P(W) = \frac{1}{2\xi}\left(1 + \frac{|W|}{\alpha \xi}\right)^{-(\alpha+1)},
    \label{eq:gdp}
\end{equation}
by maximum likelihood on the unbinned values of all grid cells.
This GDP distribution may arise from a superposition of Laplace distributions whose inverse scale follows a Gamma distribution of shape $\alpha$.
We note that this interpretation is not unique.
The Laplace distribution with exponential wings is expected for the product of two fluctuating fields such as $\bm{j}_i$ and $\bm{E}$, and the Gamma mixing accounts for variations in local energisation intensity between structures.
The scale $\xi$ sets the width, and the shape parameter $\alpha$ controls the wings, which decay as $P(W)\propto|W|^{-(\alpha+1)}$.
It is the inverse squared coefficient of variation of the Gamma-distributed rate, so small $\alpha$ indicates energy exchange concentrated in intermittent coherent structures \citep{sebastian_magnetic_2025}.
In the opposite limit, $\alpha\to\infty$, the mixing distribution collapses and \eqref{eq:gdp} reduces to a Laplace distribution of scale $\xi$, i.e.\ to the case of no intermittency.

Table~\ref{tab:budget} also gives $\sigma$ and the skewness $\mathcal{S}$ of each measured distribution, together with the fitted $\alpha$ and $\xi$.
We find that every term is narrower at $\beta_{i0}=4$, but not by the same factor.
The width decreases by a factor of $10.2$ for $W_{\mathrm{pol}}$ and $8.9$ for $W_{\mathrm{curv1}}$, against only $2.5$ for $W_{\mathrm{agy}}$.
The widest distribution therefore changes from $W_{\mathrm{pol}}$ at $\beta_{i0}=0.25$ to the diamagnetic and agyrotropic terms at $\beta_{i0}=4$.
The total decreases by a factor of $4.0$, close to the $3.9$ factor found for the volume integral.
The local statistics thus reproduce the change of composition seen in the budget.

Moreover, $\alpha$ increases for every term, suggesting that energisation becomes \emph{less} intermittent as $\beta_{i0}$ increases.
The $\beta_{i0}=4$ run has an extended sub-ion range and essentially no MHD range, whereas at $\beta_{i0}=0.25$, the injection scale lies above the ion gyroradius and the sub-ion range is short (\S\ref{sec:spectra}).
The less intermittent distributions are thus found in the run whose cascade is almost entirely kinetic.
This is consistent with the crossover, observed \textit{in situ} in turbulent space plasma and in numerical simulations, from multifractal scaling in the inertial range to non-Gaussian monoscaling below the ion gyroscale~\citep{kiyani_global_2009,leonardis_multifractal_2016,richard_turbulence_2024}.

In addition, we compute the average of $|W|$ for each term of \eqref{eq:J_dot_E}, conditioned on the local current density and normalised to its unconditional value, $\langle|W|\,|\, Q_J\rangle/\langle|W|\rangle$ (figure~\ref{fig:ws-pdfs}c).
Here $Q_J=\tfrac{1}{4}\bm{J}^2/\langle\bm{J}^2\rangle$ is the current contribution to the second invariant of the magnetic-field gradient tensor, normalised such that $\langle Q_J\rangle=1/4$, marked by the vertical dotted line.

Every term behaves the same way in both runs.
Below $\langle Q_J\rangle$, the conditional average is flat, and below unity, between $0.2$ and $0.6$ depending on the term, so the low-current volume is energised less than average.
The curves cross unity just above $\langle Q_J\rangle$ and then rise steeply, reaching a factor of $2$ to $3$ by $Q_J\simeq5$.
The parallel term rises most steeply and reaches $\simeq8$ at $\beta_{i0}=4$, so $W_\parallel$ is the term most tightly confined to the current sheets, even though it is negligible in the volume integral.
This is consistent with the formation of field-aligned electric fields in regions of electron-only reconnection~\citep{granier_electrononly_2024}, to which we return in \S\ref{sec:scales}.
The two runs are nearly indistinguishable once $Q_J$ is measured against its own mean, so this organisation is insensitive to $\beta_{i0}$.
The only systematic difference is that the $\beta_{i0}=0.25$ run reaches larger $Q_J$, its current density being more extreme relative to its mean.

Taken together with the heavy tails of figure~\ref{fig:ws-pdfs}(a,b), this locates the intermittency of the energisation.
The fits to \eqref{eq:gdp} show that energy exchange is concentrated in rare events, and the conditional averages show that these events occur in regions of intense current, where the radius of curvature also falls to the local ion gyroradius (\S\ref{sec:curvature}).
We note, however, that an enhancement of $2$ to $3$ means the strongest-current cells are preferentially energised, not that they carry most of the budget, which also depends on their filling factor.

\begin{table}
    \begin{center}
        \setlength{\tabcolsep}{2.6pt}
        \begin{tabular}{llrrrrrrrrr}
            \toprule
            & & \multicolumn{6}{c}{energisation rate \eqref{eq:J_dot_E}}
            & \multicolumn{3}{c}{pressure--strain}\\
            \cmidrule(lr){3-8}\cmidrule(lr){9-11}
            & & $W_{\mathrm{tot}}$ & $W_\parallel$ & $W_{\mathrm{dia}}$
            & $W_{\mathrm{curv1}}$ & $W_{\mathrm{agy}}$ & $W_{\mathrm{pol}}$
            & $\mathrm{PS}$ & $-p_i\theta_i$
            & $-\bm{\Pi}_i\bm{:}\bm{D}_i$\\
            \midrule
            Run A & code units
              & $30.14$ & $0.30$ & $1.58$ & $-0.80$ & $6.24$ & $22.81$
              & $11.78$ & $2.49$ & $9.29$\\
            $\beta_{i0}=0.25$ & per cent
              & $100$ & $1.0$ & $5.3$ & $-2.6$ & $20.7$ & $75.7$
              & $39.1$ & $8.3$ & $30.8$\\[3pt]
            & $\sigma\ (10^{-3})$
              & $31.5$ & $0.435$ & $28.1$ & $2.16$ & $17.7$ & $41.0$
              & \multicolumn{3}{c}{---}\\
            & $\mathcal{S}$
              & $+0.36$ & $-0.55$ & $+0.00$ & $+0.48$ & $-0.60$ & $-0.18$
              & \multicolumn{3}{c}{---}\\
            & $\alpha$
              & $7.77$ & $1.94$ & $3.30$ & $1.19$ & $3.51$ & $5.52$
              & \multicolumn{3}{c}{---}\\
            & $\xi\ (10^{-3})$
              & $18.2$ & $0.127$ & $11.8$ & $0.332$ & $7.29$ & $21.2$
              & \multicolumn{3}{c}{---}\\
            \midrule
            Run B & code units
              & $7.65$ & $-0.01$ & $3.40$ & $-0.04$ & $6.92$ & $-2.62$
              & $7.49$ & $2.29$ & $5.20$\\
            $\beta_{i0}=4$ & per cent
              & $100$ & $-0.2$ & $44.5$ & $-0.6$ & $90.5$ & $-34.2$
              & $97.9$ & $29.9$ & $68.0$\\[3pt]
            & $\sigma\ (10^{-3})$
              & $7.89$ & $0.0763$ & $8.59$ & $0.243$ & $7.10$ & $4.03$
              & \multicolumn{3}{c}{---}\\
            & $\mathcal{S}$
              & $+0.60$ & $+0.51$ & $+0.40$ & $-1.29$ & $+0.42$ & $-0.35$
              & \multicolumn{3}{c}{---}\\
            & $\alpha$
              & $8.04$ & $2.52$ & $7.29$ & $1.74$ & $9.07$ & $11.68$
              & \multicolumn{3}{c}{---}\\
            & $\xi\ (10^{-3})$
              & $4.59$ & $0.0256$ & $4.88$ & $0.0625$ & $4.27$ & $2.51$
              & \multicolumn{3}{c}{---}\\
            \bottomrule
        \end{tabular}
        \caption{Volume-integrated ion energisation rate and the statistics of its pointwise distribution for each term of \eqref{eq:J_dot_E} at peak $J_{\mathrm{rms}}$.
        For each run, the first two lines give the volume integral in code units, rounded to two decimal places, and its percentage of the total, computed from the unrounded values.
        The polarisation term is the residual \eqref{eq:Wpol_residual}, so the five terms sum to the total by construction.
        The last four lines give the standard deviation $\sigma$ and skewness $\mathcal{S}$ of the measured distribution, and the shape and scale parameters $\alpha$ and $\xi$ of the generalised double Pareto fit \eqref{eq:gdp}; $\sigma$ and $\xi$ are in units of $10^{-3}$ code units, while $\mathcal{S}$ and $\alpha$ are dimensionless.
        The last three columns give the pressure--strain interaction $\mathrm{PS} = -(\bm{P}_i\bm{\cdot}\bm{\nabla})\bm{\cdot}\bm{U}_i$ and its two parts, computed from \eqref{eq:pressure-strain}.
        They are not part of the decomposition \eqref{eq:J_dot_E}; their percentages are given relative to $W_{\mathrm{tot}}$, and they have not been fitted with \eqref{eq:gdp}.}
        \label{tab:budget}
    \end{center}
\end{table}

\subsection{Scale dependence of the decomposition}\label{sec:scales}
\begin{figure}
    \centering
    \includegraphics[width=\linewidth]{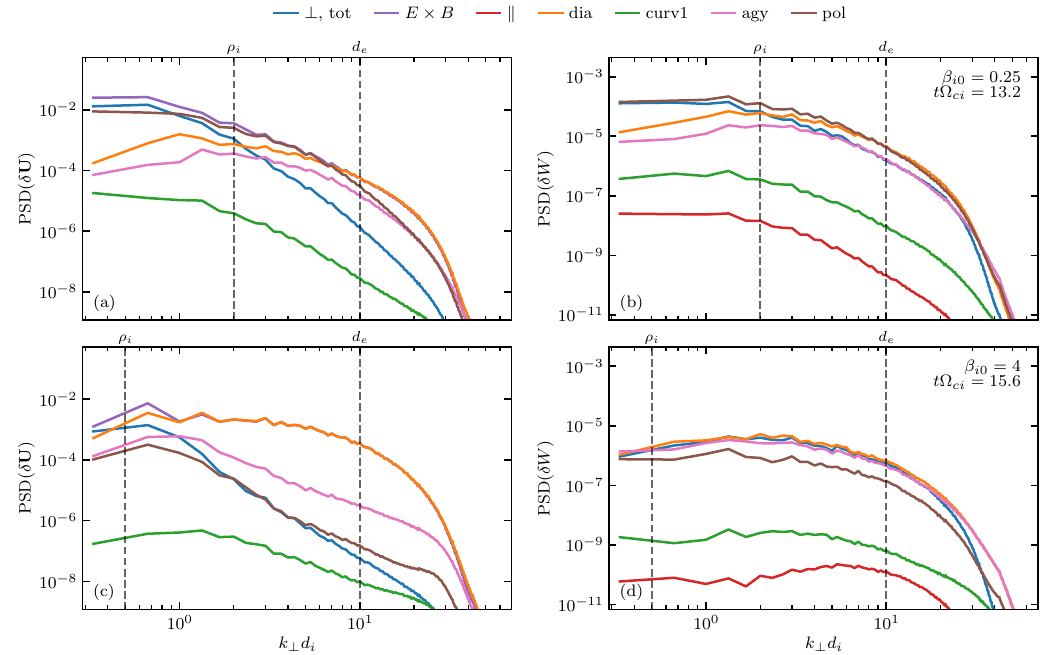}
    \caption{Reduced power spectra of (a,c) the perpendicular drift velocities and (b,d) the terms of \eqref{eq:J_dot_E}, for (a,b) run A at $t\Oci=13.20$ and (c,d) run B at $t\Oci=15.60$.
    Vertical dashed lines mark $k_\perp\rho_i=1$ and $k_\perp d_e=1$.
    All quantities are in code units.}
    \label{fig:wspectra}
\end{figure}

The spatial distributions in figures~\ref{fig:ws-beta0.25} and~\ref{fig:ws-beta4} do not show how the terms depend on scale.
Figure~\ref{fig:wspectra} therefore shows the reduced power spectra $P(k_\perp)$ of the individual drift velocities (left column) and of the individual terms of \eqref{eq:J_dot_E} (right column) for the two runs.

At $\beta_{i0}=0.25$ (figure~\ref{fig:wspectra}a,b), the $\bm{E}\times\bm{B}$ and polarisation drifts dominate the perpendicular ion flow at $k_\perp\rho_i\lesssim1$, where the diamagnetic drift is roughly two orders of magnitude smaller.
At sub-ion scales, the diamagnetic drift becomes comparable to the polarisation drift, which decreases.
The power of the agyrotropic drift is between two and three orders of magnitude below that of the $\bm{E}\times\bm{B}$ drift at $k_\perp\di<1$ and within an order of magnitude of that of the diamagnetic drift at sub-ion scales. In contrast, the curvature drift power is several orders of magnitude smaller at all scales.

In terms of energisation, the power of $W_{\mathrm{pol}}$ exceeds every other term by about an order of magnitude at $k_\perp\di<1$.
Furthermore, at sub-ion scales ($k_\perp\rho_i>1$), the diamagnetic term's power becomes comparable to that of the polarisation term, and together they dominate all other terms across all scales, with $W_\parallel$ negligible throughout the range.

At $\beta_{i0}=4$ (figure~\ref{fig:wspectra}c,d), the diamagnetic drift has a nearly flat spectrum over a broad sub-ion range and dominates the non-$\bm{E}\times\bm{B}$ flow there.
At sub-ion scales, the agyrotropic term's power becomes comparable to the diamagnetic term's, and the two track each other closely from $k_\perp\rho_i\simeq1$ down to the grid scale, whereas the polarisation term is an order of magnitude smaller than either.
In addition, the parallel-term spectrum, several orders of magnitude below the total, shows a bump around the electron skin depth.
This indicates the formation of field-aligned electric fields at these scales through electron inertia and can be attributed to the onset of electron-only reconnection, as diagnosed in the same simulation by \citet{granier_electrononly_2024}.

In both runs, the spectrum of the total is set by the largest individual term rather than by a cancellation between terms.
At $\beta_{i0}=0.25$, $\delta W_{\rm tot}^2$ lies on top of $\delta W_{\rm pol}^2$ over most of the resolved range and falls only slightly below $\delta W_{\rm pol}^2$ and $\delta W_{\rm dia}^2$ at sub-ion scales.
At $\beta_{i0}=4$, the total tracks $\delta W_{\rm dia}^2$ and $\delta W_{\rm agy}^2$ closely from $k_\perp\rho_i\simeq1$ down to $k_\perp d_e\simeq3$.
It falls below them only at the smallest resolved scales, where the compact filter acts and where every term is in any case some six orders of magnitude below its peak.

The scale-by-scale picture therefore agrees with the volume-integrated budget of table~\ref{tab:budget}.
The polarisation current carries the energisation at $\beta_{i0}=0.25$, while the diamagnetic and agyrotropic currents carry it at $\beta_{i0}=4$, at every scale and in the integral.
What changes with scale is the ordering below the dominant term, and in particular the rise of $W_{\mathrm{agy}}$ through $k_\perp\rho_i\simeq1$, which is where the ion VDF ceases to be gyrotropic.

\subsection{Time evolution}\label{sec:time}

\begin{figure}
    \centering
    \includegraphics[width=\linewidth]{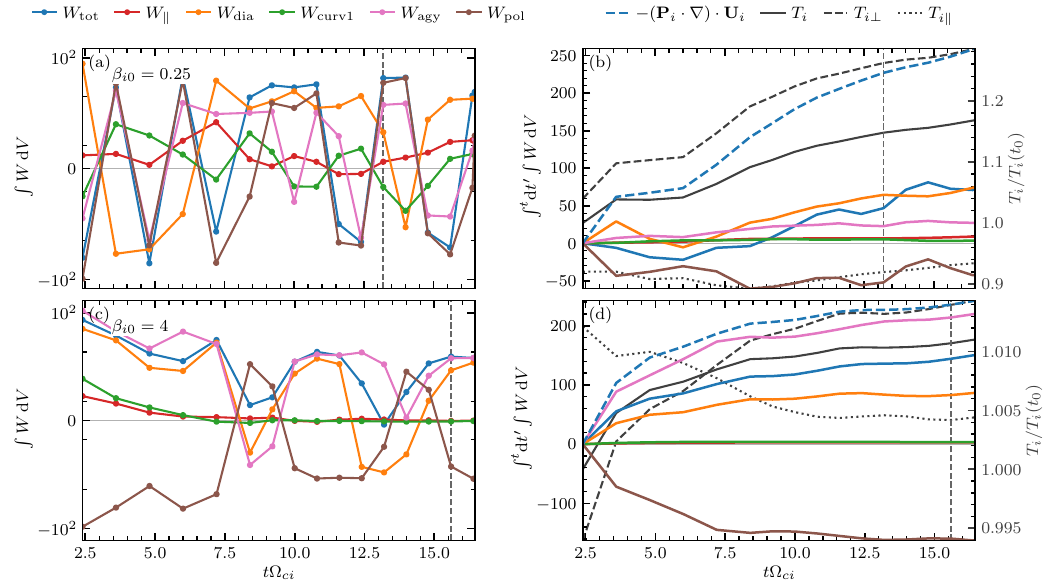}
    \caption{Volume-integrated terms of \eqref{eq:J_dot_E} as a function of time, for run A ($\beta_{i0}=0.25$, top) and run B ($\beta_{i0}=4$, bottom).
    Left, the instantaneous integral $\int W\,\dd^3 x$ on a symmetric-logarithmic ordinate.
    Right, its running time integral from $t_0\Oci=2.5$, $\int_{t_0}^t\dd t'\int W\,\dd^3 x$, i.e.\ the energy transferred by each term, together with the cumulative pressure--strain interaction (dashed blue) and the box-averaged ion temperature $T_i$ and its perpendicular and parallel components normalised to their values at $t_0\Oci=2.5$ (right-hand axis).
    The vertical dashed line marks the analysis time at which we take table~\ref{tab:budget} and the spatial maps.
    All quantities are in code units.}
    \label{fig:time_evolution}
\end{figure}

Figure~\ref{fig:time_evolution} shows the volume-integrated terms as a function of time for both runs, together with the box-averaged ion temperatures.
The earliest output times carry an initialisation transient of amplitude $\sim\!10^3$ in code units.
We attribute it to the initial state, isotropic Maxwellian ion VDFs with superposed magnetic fluctuations, not being a stationary solution of the HVM equations \citep[see e.g.][for the analogous case of sheared flows]{malara_exact_2018}.
Both the instantaneous and the cumulative panels therefore start after it, at $t\Oci=2.5$.

At $\beta_{i0}=0.25$, every term, including the total, oscillates in sign with an amplitude several times its mean, and $W_{\mathrm{tot}}$ swings between $-15$ and $+32$.
The oscillations of $W_{\mathrm{pol}}$ track those of $W_{\mathrm{tot}}$ in period and amplitude, while the remaining terms largely cancel one another.
The running time integral (figure~\ref{fig:time_evolution}b) is far less oscillatory.
The net transfer is positive and reaches $\simeq45$ code units by the analysis time.
The diamagnetic ($\simeq65$) and agyrotropic ($\simeq22$) terms carry most of it, with smaller but non-negligible contributions from the parallel electric field and the curvature term.
In contrast, the cumulative polarisation term is negative throughout ($\simeq-50$) and drives a net decrease of ion energy in the early stage of the simulation.

At $\beta_{i0}=4$, the diamagnetic, agyrotropic, and polarisation terms also oscillate with a longer period, and the total remains positive except for a single trough near $t\Oci\simeq13$, where it approaches zero.
The parallel and curvature terms contribute a little at early times and then decay to nearly zero.
The cumulative transfer, $\simeq140$ code units at the analysis time, is dominated by the agyrotropic term ($\simeq205$).
The diamagnetic term supplies roughly $40\,\%$ as much ($\simeq85$), the parallel and curvature terms are negligible, and the polarisation term is again negative throughout ($\simeq-155$).

The comparison of panels (b) and (d) reveals a result that the snapshot budget of table~\ref{tab:budget} does not show.
The cumulative transfer to the ions from $t\Oci=2.5$ to the respective analysis times is \emph{larger} at $\beta_{i0}=4$ than at $\beta_{i0}=0.25$, by a factor of about three, even though the instantaneous rate at peak $J_{\mathrm{rms}}$ is a factor of $3.9$ smaller.
We return to this in \S\ref{sec:pi-d}.

\section{Discussion}\label{sec:discussion}

\subsection{What $\bm{j}_i\bm{\cdot}\bm{E}$ does and does not measure}\label{sec:pi-d}

Before discussing the results, we clarify that the ion energisation rate in \eqref{eq:J_dot_E} is \emph{not} the ion heating rate.
Taking the dot product of the ion momentum balance \eqref{eq:momentum_balance} with $\bm{U}_i$ gives the bulk-flow energy equation,
\begin{equation}
    \dfrac{\partial \mathcal{E}_i^{\rm f}}{\partial t} + \bm{\nabla}\bm{\cdot}\left(\mathcal{E}_i^{\rm f} \bm{U}_i + \bm{P}_i\bm{\cdot}\bm{U}_i\right) = (\bm{P}_i\bm{\cdot}\bm{\nabla})\bm{\cdot}\bm{U}_i + \bm{j}_i\bm{\cdot}\bm{E},
    \label{eq:fluid-flow-energy-evol}
\end{equation}
where $\mathcal{E}_i^{\rm f} = m_i n_i |\bm{U}_i|^2 / 2$.
Subtracting it from the second moment of the Vlasov equation \eqref{eq:vlasov} gives the internal-energy equation,
\begin{equation}
    \dfrac{\partial \mathcal{E}_i^{\rm th}}{\partial t} + \bm{\nabla}\bm{\cdot}\left(\mathcal{E}_i^{\rm th} \bm{U}_i + \bm{h}_i\right) = -(\bm{P}_i\bm{\cdot}\bm{\nabla})\bm{\cdot}\bm{U}_i,
    \label{eq:internal-energy-evol}
\end{equation}
where $\mathcal{E}_i^{\rm th}= \tfrac{3}{2} n_i T_i = \tfrac{1}{2}\mathrm{Tr}(\bm{P}_i)$ and $\bm{h}_i$ is the heat-flux vector.
In addition, Poynting's theorem gives
\begin{equation}
    \dfrac{\partial \mathcal{E}^{\rm em}}{\partial t} + \bm{\nabla}\bm{\cdot}\left(\dfrac{c\,\bm{E}\times\bm{B}}{4\pi}\right) = -\sum_s \bm{j}_s\bm{\cdot}\bm{E},
    \label{eq:poynting-theorem}
\end{equation}
with $\mathcal{E}^{\rm em}= B^2/8\pi$ where the electric field energy is not included since the displacement current is neglected in the HVM model (\S\ref{sec:setup}).
The rate $\bm{j}_i\bm{\cdot}\bm{E}$ is thus the source in the ion bulk-flow equation and the ion share of the sink in the field equation.
It measures the energy the field gives up to the ions.
It is not the ion heating rate, because it says nothing about how that energy is subsequently divided between bulk flow and internal energy.
That division is governed by the pressure--strain interaction \citep{yang_energy_2017,cassak_pressure_2022,barbhuiya_pressure_2022,conley_kinetic_2024}, which transfers energy between the two reservoirs without adding to their sum.
The two diagnostics are therefore complementary.
Equation~\eqref{eq:J_dot_E} says through which current, and where, the field loses energy to the ions, whereas the pressure--strain says how much of what the ions receive ends up as heat.
A term that is large in \eqref{eq:J_dot_E} identifies the former, not the latter.

To clarify how the energy transferred to the ions is partitioned, we compute the pressure--strain interaction \citep{yang_energy_2017,cassak_pressure_2022},
\begin{equation}
    -(\bm{P}_i\bm{\cdot}\bm{\nabla})\bm{\cdot}\bm{U}_i
    = -p_i\theta_i - \bm{\Pi}_i\bm{:}\bm{D}_i ,
    \label{eq:pressure-strain}
\end{equation}
where $\theta_i = \bm{\nabla}\bm{\cdot}\bm{U}_i$ is the dilatation of the ion flow, $\bm{\Pi}_i = \bm{P}_i - p_i\bm{I}$ is the deviatoric pressure tensor, and $\bm{D}_i = \tfrac{1}{2}\bigl[\bm{\nabla}\bm{U}_i + (\bm{\nabla}\bm{U}_i)^{\mathrm{T}}\bigr] - \tfrac{1}{3}\theta_i\bm{I}$ is the traceless strain-rate tensor.
We note that $\bm{\Pi}_i$ differs from the non-gyrotropic tensor $\bm{\Pi}_i^{\text{agy}}$ of \eqref{eq:pressure_tensor_cgl}, as it also includes the gyrotropic anisotropy $p_{i\parallel}-p_{i\perp}$.
The pressure dilatation $-p_i\theta_i$ corresponds to ion heating (cooling) by compression (expansion) of the flow, whereas the Pi-D term $-\bm{\Pi}_i\bm{:}\bm{D}_i$ corresponds to heating by incompressible deformation of the flow, often referred to as ``viscous-like collisionless dissipation'' although reversible \citep{yang_energy_2017,yang_energy_2017a,bandyopadhyay_collisionallike_2023,cassak_pressure_2022}.
With the sign convention of \eqref{eq:pressure-strain}, both terms are sources of ion internal energy in \eqref{eq:internal-energy-evol}.

\begin{figure}
    \centering
    \includegraphics[width=\linewidth]{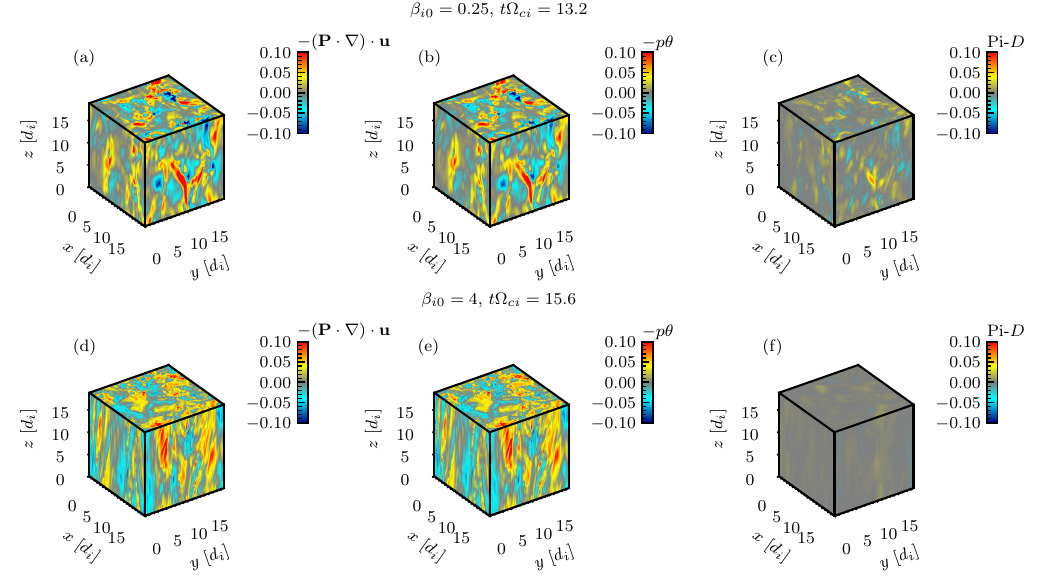}
    \caption{Ion pressure--strain interaction for run A ($\beta_{i0}=0.25$, $t\Oci=13.20$, top row) and run B ($\beta_{i0}=4$, $t\Oci=15.60$, bottom row).
    (a,d) Pressure--strain interaction $-(\bm{P}_i\bm{\cdot}\bm{\nabla})\bm{\cdot}\bm{U}_i$, (b,e) pressure dilatation $-p_i\theta_i$ and (c,f) Pi-D term $-\bm{\Pi}_i\bm{:}\bm{D}_i$, all on the faces of the simulation domain.
    The colour scale is that of figures~\ref{fig:ws-beta0.25} and~\ref{fig:ws-beta4}, and all quantities are in code units.}
    \label{fig:pstrain}
\end{figure}

Figure~\ref{fig:pstrain} shows the three terms for the two runs at the time of peak $J_{\mathrm{rms}}$, with the same colour scale as $\bm{j}_i\bm{\cdot}\bm{E}$ in figures~\ref{fig:ws-beta0.25} and~\ref{fig:ws-beta4}.
The local amplitude of the pressure--strain interaction is comparable to that of the ion energisation rate.
At $\beta_{i0}=0.25$, the pressure--strain interaction is organised in structures of a few $\di$, whereas at $\beta_{i0}=4$ it exhibits much finer structures elongated along $\bm{B}_0$.
This organisation is consistent with the finer ion-pressure structure at $\beta_{i0}=4$ (figure~\ref{fig:nonmax_highbeta}).
In both runs, pressure dilatation dominates the maps, with large local amplitudes of both signs, whereas the Pi-D term is much weaker locally, particularly at $\beta_{i0}=4$.

However, the volume-integrated values (table~\ref{tab:budget}) show the opposite ordering.
We find that the volume-integrated Pi-D term is $3.7$ and $2.3$ times larger than the pressure dilatation at $\beta_{i0}=0.25$ and $\beta_{i0}=4$, respectively.
We interpret this as pressure dilatation cancelling between compressing and expanding regions, so the Pi-D term accounts for most of the net conversion into internal energy.
This result is consistent with previous kinetic simulations of turbulence, which showed that pressure dilatation is locally large but oscillates in sign, whereas the Pi-D term dominates the net heating \citep{yang_energy_2017,pezzi_energy_2019}.

Furthermore, the pressure--strain interaction and $\bm{j}_i\bm{\cdot}\bm{E}$ together close the instantaneous ion energy budget.
Integrating \eqref{eq:fluid-flow-energy-evol} and \eqref{eq:internal-energy-evol} over the periodic domain, the flux terms vanish, and we obtain
\begin{equation}
    \dfrac{\dd}{\ddt}\int \mathcal{E}_i^{\rm th}\,\dd^3 x = -\int (\bm{P}_i\bm{\cdot}\bm{\nabla})\bm{\cdot}\bm{U}_i\,\dd^3 x,
    \label{eq:internal-energy-int}
\end{equation}
\begin{equation}
    \dfrac{\dd}{\ddt}\int \mathcal{E}_i^{\rm f}\,\dd^3 x = \int \bm{j}_i\bm{\cdot}\bm{E}\,\dd^3 x + \int (\bm{P}_i\bm{\cdot}\bm{\nabla})\bm{\cdot}\bm{U}_i\,\dd^3 x.
    \label{eq:fluid-flow-energy-int}
\end{equation}
At $\beta_{i0}=4$, we find that the pressure--strain interaction accounts for $98\,\%$ of the ion energisation rate.
This suggests that nearly all of the energy transferred from the field to the ions at this time is converted into internal energy, whereas the ion bulk-flow energy remains nearly constant, $\dd_t\!\int\mathcal{E}_i^{\rm f}\,\dd^3 x \simeq 0.16$ code units.
In contrast, at $\beta_{i0}=0.25$, the pressure--strain interaction accounts for only $39\,\%$ of $\bm{j}_i\bm{\cdot}\bm{E}$, so that the remaining $18.4$ code units increase the ion bulk-flow energy.
This value is comparable to $W_{\mathrm{pol}}=22.8$, i.e., the work done to accelerate the ion bulk flow.
We interpret this as reflecting the reversible exchange between the field and the ion bulk flow discussed later in \S\ref{sec:beta}.
We note that these are instantaneous values at a single time.
Therefore, as for the energisation, the net heating should be assessed using the time-integrated budget.

Figure~\ref{fig:time_evolution}(b,d) shows the time-integrated pressure--strain interaction together with the time evolution of the ion temperature, normalised to its value at $t\Oci=2.5$, i.e.\ after the transient phase.
First, we note that the cumulative pressure--strain interaction exceeds the cumulative ion energisation by a factor of $\simeq4.6$ at $\beta_{i0}=0.25$ and $\simeq1.7$ at $\beta_{i0}=4$ at the time of peak $J_{\mathrm{rms}}$.
Since the two differ by the change in bulk-flow energy \eqref{eq:internal-energy-int}, \eqref{eq:fluid-flow-energy-int}, the ion bulk flow loses energy over this interval in both runs, so that the internal energy gained exceeds what the field supplies.
Nevertheless, the cumulative pressure--strain interaction follows the temperature evolution remarkably well, particularly for the perpendicular temperature $T_{i\perp}$.
The ion temperature rises by $\simeq15\,\%$ in run A and by only $\simeq1\,\%$ in run B over the same interval.
However, the initial thermal energy is sixteen times larger in run B, so the absolute thermal energy gained, $\tfrac{3}{2}\Delta(n_iT_i)V$, is of the same order in both runs, $\simeq190$ and $\simeq210$ code units, respectively.

\subsection{Relation to the guiding-centre description}

Question (i) in \S\ref{sec:intro} asked whether ion energisation can be characterised when guiding-centre theory does not apply.
Equations~\eqref{eq:gc_correspondence} and~\eqref{eq:gc_residual} show that the measured budget consists of a part that regroups the betatron and Fermi terms and a part that is not captured by the guiding-centre description.
Our results show that the latter is significant.
In particular, $W_{\mathrm{agy}}$, which directly quantifies the breakdown of the guiding-centre description, accounts for $21\,\%$ of the instantaneous budget at $\beta_{i0}=0.25$ and dominates it at $\beta_{i0}=4$ even though the ion VDFs are nearly isotropic and gyrotropic in the latter case (figure~\ref{fig:nonmax_highbeta}).
Furthermore, we show that the ion energisation is preferentially enhanced in regions of large current density. 
There, the radius of curvature decreases to the local ion gyroradius, $r_c\sim\rho_i$, in the current sheets, so that the magnetic moment is not conserved.
Therefore, the guiding-centre description breaks down precisely where energisation is concentrated.

\subsection{The $\beta_i$ dependence}\label{sec:beta}

The sign of the $\beta_i$ energisation trend depends on the measure used.
At peak $J_{\mathrm{rms}}$, the volume-integrated energisation is about $3.9$ times smaller at $\beta_{i0}=4$, whereas the cumulative transfer in figure~\ref{fig:time_evolution} is about three times larger.
The two are not in conflict.
At $\beta_{i0}=0.25$, the instantaneous integral is dominated by a reversible exchange between the field and the ion bulk flow that changes sign from trough to trough, whereas the running integral measures what remains once that oscillation has averaged out.
A snapshot, even at peak $J_{\mathrm{rms}}$, may therefore report the phase of the oscillation rather than the secular transfer, and a statement about the $\beta_i$ dependence should rest on the cumulative budget (\S\ref{sec:time}).

The budget composition also changes with $\beta_{i0}$.
At $\beta_{i0}=4$, the agyrotropic term accounts for $91\,\%$ of the instantaneous budget, and the polarisation term reverses sign, so the polarisation current contributes negative work while the bulk-flow energy remains nearly constant.
However, this occurs in the run with smaller agyrotropy, $\sqrt{Q_i}\lesssim0.05$, compared with $0.42$ at $\beta_{i0}=0.25$.
To clarify this, we compute the divergence of the non-gyrotropic pressure tensor, which enters $W_{\mathrm{agy}}=\UE\bm{\cdot}\bm{\nabla}\bm{\cdot}\bm{\Pi}_i^{\text{agy}}$.
We find that $\langle|\bm{\nabla}\bm{\cdot}\bm{\Pi}_i^{\text{agy}}|\rangle = 0.066$ and $0.047$, and $|\bm{\nabla}\bm{\cdot}\bm{\Pi}_i^{\text{agy}}|_{\mathrm{rms}} = 0.082$ and $0.053$ code units, at $\beta_{i0}=0.25$ and $\beta_{i0}=4$, respectively.
Therefore, the divergence is $30$--$35\,\%$ smaller at $\beta_{i0}=4$, whereas the volume-integrated $W_{\mathrm{agy}}$ is slightly larger ($6.92$ against $6.24$ code units).
Since the $\bm{E}\times\bm{B}$ velocity is also weaker at $\beta_{i0}=4$, this suggests that the electric field and the agyrotropic drift $\bm{U}_{i,\mathrm{agy}}$ are more closely aligned at $\beta_{i0}=4$.
In other words, the large contribution of $W_{\mathrm{agy}}$ at $\beta_{i0}=4$ results from a better correlation between $\bm{E}$ and $\bm{U}_{i,\mathrm{agy}}$ rather than from a larger non-gyrotropic stress.

The reduction of $W_{\mathrm{curv1}}$ with $\beta_{i0}$ is not, as it might appear, a consequence of the reduced anisotropy ratio.
What enters \eqref{eq:J_dot_E} is the absolute anisotropy $p_{i\parallel}-p_{i\perp}$.
At fixed $B_0$ and $n_0$, the ion pressure is sixteen times larger at $\beta_{i0}=4$, so the ratio collapses even if the absolute anisotropy does not.
Indeed, we measure $p_{i\parallel}-p_{i\perp} = -0.04^{+0.04}_{-0.06}$ and $-0.02^{+0.05}_{-0.05}$ code units in runs~A and~B, respectively, which are comparable, whereas the spread of $p_{i\perp}/p_{i\parallel}$ differs by an order of magnitude.
The $\beta_i$ dependence of $W_{\mathrm{curv1}}$ is instead carried by $\UE\bm{\cdot}\bm{\kappa}$, whose rms falls by a factor of six between the runs, and within that partly by the curvature, whose rms falls by a factor of three, from $0.1\,\di^{-1}$ to $0.03\,\di^{-1}$, with the normalised distribution unchanged.
The measured values satisfy $|\bm{\kappa}|_{\mathrm{rms}}\rho_i \simeq 0.05$ in both runs, i.e., \ $|\bm{\kappa}|_{\mathrm{rms}}\propto\rho_i^{-1}\propto\beta_{i0}^{-1/2}$.
Its origin remains to be established.
The volume integral of $W_{\mathrm{curv1}}$ falls by more than the rms of $\UE\bm{\cdot}\bm{\kappa}$, from $-0.80$ to $-0.04$.
Since it is the signed integral of a quantity that changes sign throughout the volume, the additional reduction reflects a loss of correlation between the anisotropy and $\UE\bm{\cdot}\bm{\kappa}$ rather than a further fall in amplitude.

\subsection{Limitations}

The turbulence studied here is injected in the three lowest wavenumber shells of the box, so there is no extended MHD cascade above the ion scales.
The results apply to magnetosheath-like turbulence rather than to the solar wind.
The injection band, $0.3\le k\di\le1$, is the same in both runs, but $\rho_i$ is not.
It corresponds to $k\rho_i\simeq0.15$--$0.5$ at $\beta_{i0}=0.25$ and $0.6$--$2$ at $\beta_{i0}=4$, so run~B has essentially no range of scales above the ion gyroradius and its turbulence is kinetic from the injection scale down.
Comparisons between the runs at fixed $k\di$ therefore sample different parts of the cascade.
The mass ratio is reduced to $m_i/m_e=100$, and the grid spacing is $0.74\,d_e$, so the sub-$d_e$ range is unresolved and the sub-ion range at $\beta_{i0}=0.25$ spans less than a decade.
Turbulence decays rather than being driven, so we measure the budget against a moving background.
Finally, we obtain the polarisation term as the residual \eqref{eq:Wpol_residual}, which includes numerical errors and the effect of the numerical filter, and is therefore the least independently constrained quantity in the budget.

\section{Conclusions}\label{sec:conclusions}
We presented an analysis of the Eulerian ion energisation rate in three-dimensional hybrid-Vlasov--Maxwell simulations of decaying turbulence at $\beta_{i0}=0.25$ and $\beta_{i0}=4$, based on its decomposition into the work done on the parallel, diamagnetic, curvature, agyrotropic and polarisation ion currents.
Our main results are as follows.

\begin{enumerate}
    \item \textbf{The Eulerian decomposition is exact and differs from the guiding-centre description, and the magnetic moment is expected not to be conserved where the energisation happens.}
        Equation~\eqref{eq:J_dot_E} requires neither gyrotropy nor scale separation.
        By comparing it with the guiding-centre energisation rate \eqref{eq:gc_correspondence}, we show that the curvature term does not correspond to the Fermi term.
        Therefore, a small $W_{\mathrm{curv1}}$ does not imply weak Fermi energisation.
        In addition, the difference between the two descriptions \eqref{eq:gc_residual} reduces to a few terms, including the agyrotropic term, which has no guiding-centre counterpart.
        The radius of curvature falls to the local ion gyroradius in the current sheets of both runs.

    \item \textbf{Which current carries the energisation changes with $\beta_{i0}$, and the explicitly non-gyrotropic term is never negligible across and below the ion scales.}
        The polarisation current carries the budget at $\beta_{i0}=0.25$, and the diamagnetic and agyrotropic currents carry it at $\beta_{i0}=4$.
        $W_{\mathrm{agy}}$ carries $21\,\%$ of the instantaneous budget at $\beta_{i0}=0.25$ and $91\,\%$ at $\beta_{i0}=4$, and it dominates the cumulative budget at $\beta_{i0}=4$.

    \item \textbf{The local and volume-integrated pictures agree.}
        The same currents carry the energisation at every resolved scale and in the volume integral, with the total spectrum following the largest single term rather than resulting from a cancellation between terms.

    \item \textbf{The instantaneous budget is dominated by a reversible exchange, and the cumulative budget is the relevant measure.}
        The instantaneous measure is dominated by a reversible exchange that oscillates in sign.
        The instantaneous rate at peak $J_{\mathrm{rms}}$ is $3.9$ times smaller at $\beta_{i0}=4$, whereas the cumulative transfer is larger.
        Statements about the $\beta_i$ dependence should therefore rest on the cumulative budget.

    \item \textbf{The ion VDFs are not held at marginal stability, and the fall of $W_{\mathrm{curv1}}$ with $\beta_{i0}$ is carried by the curvature, not by the anisotropy.}
        At $\beta_{i0}=0.25$ the mirror threshold bounds the upper envelope of the anisotropy distribution but not its bulk.
        At $\beta_{i0}=4$ the whole distribution lies well inside every threshold.
        The absolute anisotropy $p_{i\parallel}-p_{i\perp}$ is comparable in the two runs, so that the collapse of $p_{i\perp}/p_{i\parallel}$ with $\beta_{i0}$ follows the sixteen-fold rise of the ion pressure.
        The adiabaticity parameter $r_c/\rho_i$ is essentially independent of $\beta_{i0}$, while the absolute curvature scales as $\rho_i^{-1}$.
        The $\beta_{i0}$ dependence of $W_{\mathrm{curv1}}$ follows $\UE\bm{\cdot}\bm{\kappa}$, whose rms falls in step with $|\bm{\kappa}|_{\mathrm{rms}}$ and with the ion velocity fluctuations.
\end{enumerate}

Finally, $\bm{j}_i\bm{\cdot}\bm{E}$ is the rate at which the electromagnetic field gives energy to the ions, not the rate at which the ions are heated.
How much of it ends as internal energy is set by the pressure--strain interaction \eqref{eq:pressure-strain}, and the two diagnostics are needed together.
All quantities entering \eqref{eq:J_dot_E}, except the time derivative in $W_{\mathrm{pol}}$, are accessible to multi-spacecraft measurements in the terrestrial magnetosheath, where these results can be tested directly~\citep{bandyopadhyay_statistics_2020,yang_quantifying_2023,richard_nonmaxwellianity_2025}.

\section*{Acknowledgements}
L.R. thanks Archie F. A. Bott and Benjamin D. G. Chandran for insightful discussions.

\section*{Funding}
L.R. has received funding from the European Union's Horizon Europe research and innovation programme under the Marie Skłodowska-Curie grant agreement No.~101272207 (``MicroDynaMIQ'').
Funded by the European Union.
Views and opinions expressed are, however, those of the authors only and do not necessarily reflect those of the European Union or the European Research Executive Agency.
Neither the European Union nor the granting authority can be held responsible for them.
S.S.C. is supported by the National Research Agency (ANR) grant ``MiCRO'' with the reference number ANR-23-CE31-0016.
O.P. is supported by the FIS2 Starting Grant FIS-2023-00246 ``PhAse-sPAce cOmplexity in turbulent nearly-reversible plasmas (PAPAO)'' (CUP B53C24009610001) funded by the Italian Ministry of University and Research.
J.M.T. was supported by NASA Grant 80NSSC23K0099.

\section*{Declaration of generative AI use}
L.R. acknowledges the use of Claude Opus 5.5 (Anthropic) for assistance with post-processing and writing the manuscript.
All AI-generated content was reviewed and verified by the authors, who take full responsibility for the content of the publication.

\section*{Declaration of interests}
The authors report no conflict of interest.

\appendix

\section{Normalisation of the energy-conversion equations}\label{app:norm}

The HVM code uses the normalisation of \citet{valentini_hybridvlasov_2007},
\begin{gather}
    \bar{U} = \vA, \quad \bar{\omega} = \Oci, \quad
    \bar{l} = \vA/\Oci = c/\omega_{pi} = \di, \quad
    \bar{P} = \bar{n} m_i \vA^2, \\
    \bar{E} = \bar{m} \vA \Oci / e, \quad \bar{B} = \bar{m} c \Oci / e, \quad
    \bar{q} = e, \quad \bar{m} = m_i,
\end{gather}
where $\vA$ is the Alfvén speed, $\Oci$ the ion cyclotron frequency, $e$ the elementary charge, $m_i$ the ion mass and $\bar{n}$ a reference density.
In these units, the three normalisation ratios that appear in \eqref{eq:Uperp} all reduce to the Alfvén speed,
\begin{equation}
    \dfrac{c\bar{E}}{\bar{B}} = \vA, \qquad
    \dfrac{c\bar{P}}{\bar{q}\,\bar{n}\,\bar{B}\,\bar{l}}
      = \dfrac{\vA^2}{\Oci\,\di} = \vA, \qquad
    \dfrac{\bar{m}\,c\,\bar{U}\,\bar{\omega}}{\bar{q}\,\bar{B}} = \bar{U} = \vA ,
\end{equation}
so that the normalised drift decomposition is identical in form to \eqref{eq:Uperp} with all factors of $c$ removed,
\begin{multline}
    \tilde{\bm{U}}_{i\perp}
    = \dfrac{\tilde{\bm{E}}\times\tilde{\bm{B}}}{\tilde{B}^2}
    + \dfrac{\tilde{\bm{B}}\times\tilde{\bm{\nabla}}\tilde{p}_{i\perp}}
           {\,\tilde{n}_i\,\tilde{B}^2}
    + \dfrac{\tilde{p}_{i\parallel}-\tilde{p}_{i\perp}}{\,\tilde{n}_i\,\tilde{B}^2}\,
        \tilde{\bm{B}}\times\tilde{\bm{\kappa}} \\
    + \dfrac{\tilde{\bm{B}}\times\tilde{\bm{\nabla}}\bm{\cdot}\tilde{\bm{\Pi}}_i^{\text{agy}}}
           {\,\tilde{n}_i\,\tilde{B}^2}
    + \dfrac{\tilde{m}_i}{\,\tilde{B}^2}\,
        \tilde{\bm{B}}\times\dfrac{\dd\tilde{\bm{U}}_i}{\dd\tilde{t}},
    \label{eq:Uperp_norm}
\end{multline}
and \eqref{eq:J_dot_E} becomes
\begin{equation}
    \tilde{\bm{j}}_{i}\bm{\cdot}\tilde{\bm{E}}
    = \tilde{j}_{i\parallel} \tilde{E}_{\parallel}
    + \tilde{\UE}\bm{\cdot} \tilde{\bm{\nabla}}\tilde{p}_{i\perp}
    + \bigl(\tilde{p}_{i\parallel}-\tilde{p}_{i\perp}\bigr)\,
        \tilde{\UE}\bm{\cdot}\tilde{\bm{\kappa}}
    + \tilde{\UE}\bm{\cdot}\tilde{\bm{\nabla}}\bm{\cdot}\tilde{\bm{\Pi}}_i^{\text{agy}}
    + \tilde{m}_i \tilde{n}_i\tilde{\UE}\bm{\cdot}\dfrac{\dd\tilde{\bm{U}}_i}{\dd\tilde{t}},
    \label{eq:Js.E_norm}
\end{equation}
with $\tilde{\UE} = \tilde{\bm{E}}\times\tilde{\bm{B}}/\tilde{B}^2$.

\section{Derivation of the guiding-centre correspondence}\label{app:gc}

This appendix derives \eqref{eq:gc_correspondence} and \eqref{eq:gc_residual}.
Apart from \eqref{eq:gc_energisation} itself, every step is an identity.

\subsection{Field strength and direction}\label{app:gc_identities}

Since $(\bm{E}\times\bm{B})\times\bm{B} = B^2\bigl(\bhat E_\parallel - \bm{E}\bigr)$, the electric field can be written exactly as $\bm{E} = E_\parallel\bhat - \UE\times\bm{B}/c$.
We note that this is an identity rather than a frozen-in assumption, since any Hall or electron-pressure contribution to $\bm{E}_\perp$ is contained in $\UE$.
Faraday's law therefore splits, without approximation, as
\begin{equation}
    \DE\bm{B}
    = \bigl(\bm{B}\bm{\cdot}\bm{\nabla}\bigr)\UE
      - \bm{B}\,\bm{\nabla}\bm{\cdot}\UE
      - c\,\bm{\nabla}\times\bigl(E_\parallel\bhat\bigr).
    \label{eq:faraday-gc}
\end{equation}
We project \eqref{eq:faraday-gc} on $\bhat$ and use $\bhat\bm{\cdot}\DE(B\bhat)=\DE B$ together with two identities.
Applying $\bhat\bm{\cdot}\bm{\nabla}$ to $\bhat\bm{\cdot}\UE=0$ gives $\bhat\bm{\cdot}(\bhat\bm{\cdot}\bm{\nabla})\UE = -\UE\bm{\cdot}\bm{\kappa}$, and Ampère's law without displacement current gives $\bhat\bm{\cdot}(\bm{\nabla}\times\bhat) = 4\pi J_\parallel/cB$.
The field strength then evolves as
\begin{equation}
    \DE \ln B
    = -\,\bm{\nabla}\bm{\cdot}\UE
      - \UE\bm{\cdot}\bm{\kappa}
      - \dfrac{4\pi J_\parallel E_\parallel}{B^2},
    \label{eq:dlnBdt}
\end{equation}
where $J_\parallel$ is the \emph{total} parallel current, which at sub-ion scales is carried mostly by the electrons~\citep{franci_solar_2018,cerri_kinetic_2019}.
Thus, the field strength follows the $\bm{E}\times\bm{B}$ flow except where a parallel electric field acts along twisted field lines.
Projecting \eqref{eq:faraday-gc} instead on the plane perpendicular to $\bhat$ and contracting with $\UE$ gives
\begin{equation}
    \UE\bm{\cdot}\DE\bhat
    = \tfrac{1}{2}\,\partial_\parallel|\UE|^2
      - \dfrac{c}{B}\,\UE\bm{\cdot}
        \bigl[\bm{\nabla}\times\bigl(E_\parallel\bhat\bigr)\bigr],
    \label{eq:dbhatdt}
\end{equation}
so that $W_{\mathrm{rot}}$ can be evaluated without a time derivative.

\subsection{The guiding-centre rate}\label{app:gc_rate}

The non-relativistic guiding-centre energy equation of \citet{northrop_adiabatic_1963}, for a particle of parallel velocity $v_\parallel$ and magnetic moment $\mu$, may be written
\begin{equation}
    \dfrac{\dd \varepsilon}{\ddt}
    = q v_\parallel E_\parallel
    + \mu\,\DE B
    + m v_\parallel^2\,\UE\bm{\cdot}\bm{\kappa}
    + m v_\parallel\,\UE\bm{\cdot}\DE\bhat
    + \dfrac{m}{2}\,\dfrac{\dd|\UE|^2}{\ddt},
    \label{eq:northrop}
\end{equation}
with $\dd/\ddt$ following the guiding centre.
The third and fourth terms both arise from $\bhat\bm{\cdot}\dd\UE/\ddt = -\UE\bm{\cdot}\dd\bhat/\ddt$ in the parallel equation of motion, with $\dd\bhat/\ddt = \DE\bhat + v_\parallel\bm{\kappa}$.
The third is the centrifugal term and the fourth the work done as the field direction rotates following the $\bm{E}\times\bm{B}$ flow.
Averaging \eqref{eq:northrop} over the guiding centres in a local volume, using $n\langle\mu\rangle B=p_{i\perp}$ and $mn\langle v_\parallel^2\rangle = p_{i\parallel}+mnU_{i\parallel}^2$, and dropping the fourth term as in \citet{dahlin_mechanisms_2014} and \citet{li_particle_2017}, gives \eqref{eq:gc_energisation}.
% NOTE: please verify (B.4) against Northrop (1963) before submission.

\subsection{Correspondence and residual}\label{app:gc_residual}

Substituting \eqref{eq:dlnBdt} into \eqref{eq:gc_energisation} and using $W_{\mathrm{dia}} = \bm{\nabla}\bm{\cdot}(p_{i\perp}\UE) - p_{i\perp}\bm{\nabla}\bm{\cdot}\UE$, the two occurrences of $p_{i\perp}\UE\bm{\cdot}\bm{\kappa}$ cancel and \eqref{eq:gc_correspondence} follows.
Subtracting it from \eqref{eq:J_dot_E} leaves $W_{\mathrm{agy}} + W_{\mathrm{pol}} - W_{\mathrm{pol}}^{\rm LGC}$ plus the flux and the twist.
To evaluate $W_{\mathrm{pol}} - W_{\mathrm{pol}}^{\rm LGC}$, we split the ion bulk velocity exactly as $\bm{U}_i = \UE + U_{i\parallel}\bhat + \uperp$, where $\uperp$ is the sum of the diamagnetic, curvature, agyrotropic and polarisation drifts of \eqref{eq:Uperp}, and write $\Di \equiv \partial_t + \bm{U}_i\bm{\cdot}\bm{\nabla}$ for the derivative in $W_{\mathrm{pol}}$.
Inserting the split into $W_{\mathrm{pol}} = m_in_i\UE\bm{\cdot}\Di\bm{U}_i$ and expanding $\Di\bhat = \DE\bhat + U_{i\parallel}\bm{\kappa} + (\uperp\bm{\cdot}\bm{\nabla})\bhat$, the term $m_in_iU_{i\parallel}^2\UE\bm{\cdot}\bm{\kappa}$ cancels the centrifugal part of $W_{\mathrm{pol}}^{\rm LGC}$ identically, and $\tfrac{1}{2}m_in_i(\Di-\Dg)|\UE|^2 = m_in_i\UE\bm{\cdot}(\uperp\bm{\cdot}\bm{\nabla})\UE$.
The remaining terms are $W_{\mathrm{rot}}$ and
\begin{equation}
    W_\perp = m_i n_i\Bigl[
        U_{i\parallel}\,\UE\bm{\cdot}
          \bigl(\uperp\bm{\cdot}\bm{\nabla}\bigr)\bhat
        + \UE\bm{\cdot}\Di\uperp
        + \UE\bm{\cdot}\bigl(\uperp\bm{\cdot}\bm{\nabla}\bigr)\UE
        \Bigr],
    \label{eq:Wperp}
\end{equation}
which vanishes identically if the perpendicular ion flow is purely $\bm{E}\times\bm{B}$.
This gives \eqref{eq:gc_residual}.

\bibliographystyle{jpp}
\bibliography{main}

\end{document}